\documentclass[prx,reprint,groupedaddress,superscriptaddress]{revtex4-2}
\usepackage[resetlabels]{multibib}
\usepackage{lipsum}
\usepackage{graphicx}
\usepackage{dcolumn}
\usepackage{bm}
\usepackage{amsmath,amssymb,amstext,mathtools}
\usepackage{physics}
\usepackage{siunitx}
\usepackage{csquotes}
\usepackage[svgnames]{xcolor}
\usepackage[colorlinks=true,
            linkcolor=red,
            urlcolor=blue,
            citecolor=DarkGreen]{hyperref}
\usepackage[capitalize]{cleveref}
\usepackage{natbib}
\usepackage{dsfont}

\usepackage{etoolbox}

\makeatletter

\renewcommand\section{\@startsection
  {section}           
  {1}                 
  {\z@}               
  {-3.5ex \@plus -1ex \@minus -.2ex} 
  {.01ex \@plus .0ex}               
  {\normalfont\fontsize{10}{11}\bfseries\raggedright}}  

\renewcommand\subsection{\@startsection
  {subsection}
  {2}
  {\z@}
  {0ex\@plus 0ex \@minus 0ex}
  {0.ex \@plus .1ex}  
  {\normalfont\fontsize{10}{11}\bfseries\raggedright}}

\makeatother

\renewcommand{\selectlanguage}[1]{}

\crefname{appendix}{Methods}{Methods}
\Crefname{appendix}{Methods}{Methods}

\begin{document}

\title{Artificial discovery of lattice models for wave transport}

\author{Jonas~Landgraf}\email{Jonas.Landgraf@mpl.mpg.de}
\affiliation{Max Planck Institute for the Science of Light, Staudtstr.~2, 91058 Erlangen, Germany}
\affiliation{Physics Department, University of Erlangen-Nuremberg, Staudstr.~5, 91058 Erlangen, Germany}
\author{Clara~C.~Wanjura}
\affiliation{Max Planck Institute for the Science of Light, Staudtstr.~2, 91058 Erlangen, Germany}
\author{Vittorio~Peano}
\affiliation{Max Planck Institute for the Science of Light, Staudtstr.~2, 91058 Erlangen, Germany}
\author{Florian~Marquardt}
\affiliation{Max Planck Institute for the Science of Light, Staudtstr.~2, 91058 Erlangen, Germany}
\affiliation{Physics Department, University of Erlangen-Nuremberg, Staudstr.~5, 91058 Erlangen, Germany}

\date{\today}


\begin{abstract}
    Wave transport devices, such as amplifiers, frequency converters, and nonreciprocal devices, are essential for modern communication, signal processing, and sensing applications. 
    Of particular interest are traveling wave setups, which offer excellent gain and bandwidth properties. So far, the conceptual design of those devices has relied on human ingenuity. This makes it difficult and time-consuming to explore the full design space under a variety of constraints and target functionalities. 
    In our work, we present a method that automates this challenge. By optimizing the discrete and continuous parameters of periodic coupled-mode lattices, our approach identifies the simplest lattices that achieve the target transport functionality, and we apply it to discover new schemes for directional amplifiers, isolators, and frequency demultiplexers. Leveraging automated symbolic regression tools, we find closed analytical expressions that facilitate the discovery of generalizable construction rules. Moreover, we utilize important conceptual connections between the device transport properties and non-Hermitian topology. The resulting structures can be implemented on a variety of platforms, including microwave, optical, and optomechanical systems. Our approach opens the door to extensions like the artificial discovery of lattice models with desired properties in higher dimensions or with nonlinear interactions. 
\end{abstract}

\maketitle

Designing lattice models with desired target properties is a common challenge in physics.  Often, the motivation is to discover the simplest model that implements an interesting emergent behavior, be it superconductivity with the Hubbard model~\cite{Micnas_1990},  topologically robust transport with the Haldane model~\cite{Haldane1998}, and anyonic excitations for fault-tolerant quantum computing with the Kitaev model~\cite{Kitaev_2006}. Another common motivation is to implement a desired wave transport behavior in the design of devices such as filters \cite{yariv1973,Pozar_microwave_engineering,naaman_synthesis_2022}, and traveling wave amplifiers~\cite{ho2012wideband,macklin2015near,white2015traveling, Peano2016, Mittal_2018, McDonald2018, wanjura2020topological}. 


So far, the design process of lattice models or, more generally, periodic structures has either relied on human ingenuity or black‑box inverse design methods~\cite{molesky2018inverse}. A major challenge and opportunity is to devise automated methods that not only produce solutions but also conceptual understanding, an aspiration central to the field of artificial scientific discovery~\cite{lindsay1993dendral,king2009automation,lennon2019efficiently,moon2020machine,duris2020bayesian}. 
Benefiting from the rapid development of machine learning and artificial intelligence, this emerging field aims 
to automate all aspects of scientific research with a focus on interpretability and the generation of new conceptual insights. 
An important application of this approach is the automated design of experimental setups, for example, optical setups for the generation of entangled quantum states~\cite{krenn2016automated,krenn_conceptual_2021}, components for superconducting quantum computers \cite{menke_automated_2021}, or scattering devices based on coupled-mode systems \cite{AutoScatter}.
In this setting, a clever representation of the experimental setups is an important stepping stone to generating conceptual understanding automatically. In particular, representing the discovered setups as graphs provides an intuitive visualization of their structure, enabling human scientists to recognize patterns and propose generalizations \cite{krenn_conceptual_2021,AutoScatter}.  
Another fruitful approach mimics human scientists in their focus on ideal solutions that exactly fulfill a set of target properties. 
By combining this concept with a clever search strategy~\cite{AutoScatter}, one can efficiently identify an 
exhaustive list of all ideal, irreducible solutions, i.e., solutions that cannot be further simplified. This allows one to postselect the solutions most suitable for the considered hardware platform after the search is already concluded \cite{AutoScatter}. 



In this work, we develop an artificial scientific discovery method combining these ideas to design lattice models with desired target properties. More specifically, we focus on one-dimensional models implementing desired wave scattering functionalities.  Here, we represent the scattering setups as lattice models with a varying number of sublattices and connectivity. Our method optimizes both the discrete structure and the continuous parameters. By using a transfer-matrix approach, it identifies the setups that implement the ideal transport behavior in the asymptotic limit of an infinite lattice. 
Using automated symbolic regression~\cite{schmidt2009distilling,brunton2016discovering,udrescu2020ai}, we find closed expressions relating the transport properties to the parameters of the discovered lattice models, which enables the discovery of general construction rules. 
We utilize our approach to discover new schemes for directional amplifiers with optimized gain and bandwidth, isolators with enhanced bandwidth, and frequency demultiplexers that selectively amplify signals within different frequency ranges in different directions. Our solutions are transferable and can be implemented on several platforms, including plasmonic waveguides~\cite{Wetter2023Observation}, photonic crystal nanobeams~\cite{slim2024optomechanical}, superconducting circuits~\cite{macklin2015near}, and hybrid platforms, such as optomechanical superconducting circuits~\cite{youssefi2022topological}.

\begin{figure*}[t]
    \centering
    \includegraphics[width=\linewidth]{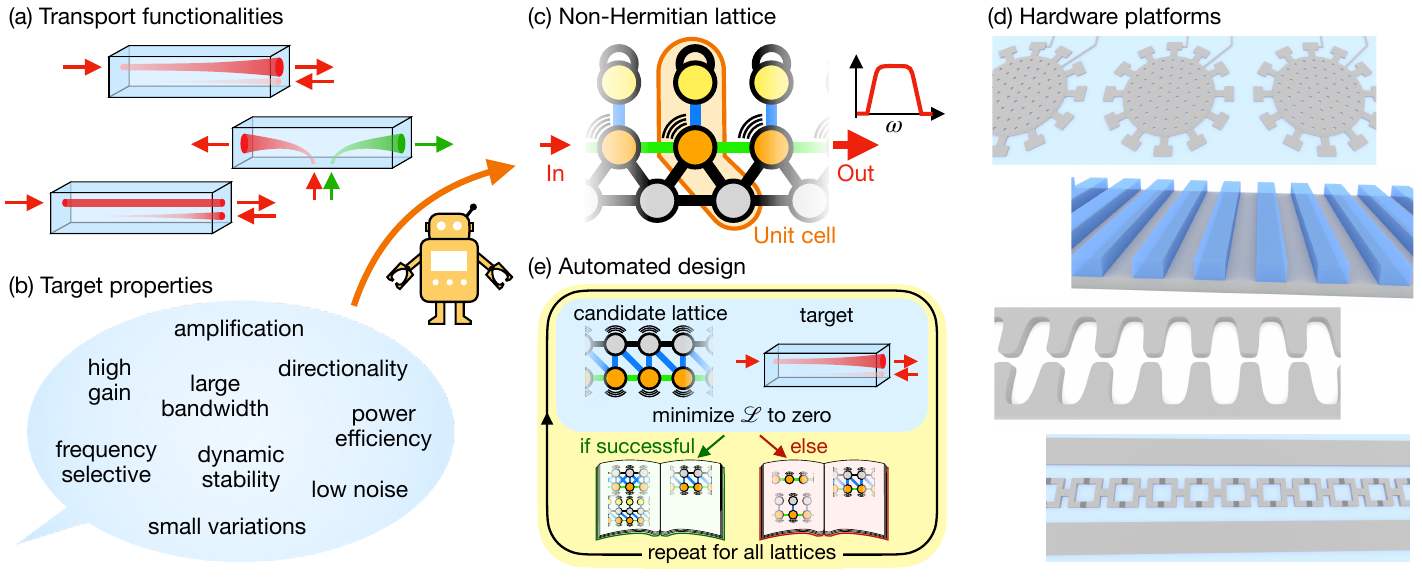}
    \caption{
    \textbf{Automated discovery of lattice models for optimized wave transport.}
    (a) Possible target functionalities for wave transport. These include unidirectional amplifiers, frequency demultiplexers, and isolators (from top to bottom).
    (b) Target properties relevant for the design of wave transport devices.
    (c) Non-Hermitian lattice model. We consider an open one-dimensional lattice of identical unit cells, each of which consists of multiple modes (differently colored circles). 
    The modes can be coupled via two-mode squeezing (blue edges), and real and complex-valued beamsplitter interactions (black and green edges). They can be detuned (black self-loops) and coupled to external waveguides or baths, leading to dissipation (represented by waves emanating from the mode). 
    Such a lattice can amplify an input signal from one end to the other. The amplification is characterized by the gain rate, which is frequency dependent (see inset).
    (d) Illustrations of some of the many hardware platforms suitable for wave transport in periodic structures. These include superconducting optomechanical circuits \cite{youssefi2022topological}, plasmonic waveguides \cite{Wetter2023Observation}, photonic crystal nanobeams \cite{slim2024optomechanical}, and superconducting circuits \cite{macklin2015near} (from top to bottom).
    (e) Optimization protocol. Given a candidate lattice, the continuous optimization (blue background) optimizes the continuous parameters to achieve the target functionality defined over the loss function $\mathcal{L}$. If the loss can be minimized to zero, the lattice is added to the list of valid lattices (green book); otherwise, to the list of invalid lattices (red book). The discrete optimization (yellow background) suggests new lattices until all possible lattices have been sorted into the two lists.
    }
    \label{fig:conceptual_figure}
\end{figure*}

Compared to previous work that was limited to setups with a small number of modes~\cite{AutoScatter}, our lattice-based method exploits periodicity to design traveling-wave setups with suitable band structures. 
This overcomes the bandwidth limitations of few-mode systems \cite{ho2012wideband}, such as the limited gain-bandwidth product in the context of amplification, while keeping the design and fabrication complexity low. Our work also ties into recent works relating directional amplification to non-Hermitian topology \cite{McDonald2018,Porras2018,wanjura2020topological,slim2024optomechanical,Busnaina2024Quantum}. 
We show that our transfer-matrix approach offers a new connection between those concepts and 
unifies existing approaches for treating non-Hermitian topological systems.

\begingroup
\let\MakeTextUppercase\relax
\section*{Results}
\endgroup
\subsection*{Open lattice model. }

We consider an open one-dimensional lattice of $N$ identical unit cells, see \cref{fig:conceptual_figure}(c). Each unit cell contains $M$ bosonic modes, described by their ladder operators $\hat{a}_{\mathbf{j}=j,m}$,  
where $j=1,\ldots N$ labels the unit cell and $m=1,\ldots, M$ the sublattice.  The modes can decay into an intrinsic loss channel (represented by a wave emanating from the mode).  Modes in the outermost unit cells are coupled to the input-output ports of the device (represented by arrows), also leading to decay.
We denote the decay rate as $\kappa_{m}$, and the fields entering and exiting the device as $a_{\mathbf{j}, {\rm in}}$ and $a_{\mathbf{j}, {\rm out}}$, respectively. 




This framework can describe a wide range of systems, including electrical, mechanical, photonic, superconducting, or hybrid lattices, see \cref{fig:conceptual_figure}(d). Such systems are weakly nonlinear and can be driven by multiple coherent drives, giving rise to effective quadratic couplings between the modes. Such quadratic couplings are of two types. On the one hand, two-mode squeezing interactions,  $\nu^{\rm (j-j')}_{mm'} \hat{a}_{\mathbf{j}}^\dag \hat{a}_\mathbf{j'}^\dag+\mathrm{H.c.}$ (represented as blue edges), create or destroy entangled pairs of excitations and are responsible for amplifying incoming signals. 
On the other hand, beamsplitter couplings, $g^{\rm (j-j')}_{mm'} \hat{a}_{\mathbf{j}}^\dag \hat{a}_{\mathbf{j'}}+\mathrm{H.c.}$, exchange excitations between two modes. Beamsplitter couplings can be laser-mediated or passive, e.g., induced by the overlap between geometrically adjacent modes. In the latter case, they can be described by real coupling constants $g$.
Laser-mediated couplings also enable complex coupling rates. 
As a result, an excitation moving on a closed loop in the lattice can accumulate a phase, which acts like an artificial magnetic flux \cite{fang_generalized_2017}. 
To differentiate these two cases, 
we distinguish between complex-valued beamsplitter couplings with $g\!\in\!\mathbb{C}$ (green edges) and real-valued beamsplitter couplings $g\!\in\!\mathbb{R}$ (black edges) in our graphical representation. 

We note that the effective time-independent quadratic couplings described above are defined in a rotating frame where each mode rotates at the appropriate reference frequency. The reference frequency is the same for passively coupled modes, but it can differ for two modes if a laser mediates their coupling, thereby introducing frequency conversion. 
Below, we denote by $\omega$ the frequency offset from the reference frequency, and by $\Delta_m$ the detuning of the mode $\mathbf{j}=j,m$ from its respective reference frequency (illustrated as a self-loop in \cref{fig:conceptual_figure}(c)). \\

\subsection*{Optimization scheme. } 
To design lattices with the desired functionalities, our method must 
search through the discrete search space of possible mode connectivities. Simultaneously, it has to find suitable values for the lattice's continuous parameters, i.e., the couplings and loss rates. 
Inspired by the automated design method for few-mode setups presented in \cite{github_autoscattering,AutoScatter}, our approach for periodic structures performs a two-step procedure: First, a discrete optimization suggests a new candidate lattice (yellow box in \cref{fig:conceptual_figure}(e)), and an embedded continuous optimization (blue box) optimizes the values of the associated continuous parameters. If the continuous optimization was successful, the candidate lattice is added to the list of valid lattices (green book); otherwise, to the list of invalid lattices (red book). The discrete optimization performs an exhaustive search over all possible lattices, up to a certain complexity. It ends up with a list of the simplest valid lattices, so-called {\it irreducible} lattices. These are the lattices that cannot be further simplified by setting any coupling rate or phase to zero without becoming invalid.


Similar to \cite{github_autoscattering,AutoScatter}, we discover setups with target functionalities by optimizing the scattering matrix $S$, i.e., the transmission coefficients $S_{\mathbf{j}\mathbf{j}'}$ between pairs of input and output ports,  $\mathbf{j}'$ and $\mathbf{j}$, respectively. 
Here, we go beyond this work and automatically discover lattice models enabling us to overcome the bandwidth limitations of few-mode systems \cite{ho2012wideband} and to impose the desired target behavior over a larger bandwidth.
To achieve this, we optimize directly the asymptotic properties of the transmission $S_{\mathbf{j}\mathbf{j}'}(\omega)$ between distant ports, $|j-j'|\gg 1$. The asymptotic expressions then can be used to infer asymptotically exact approximations of the transmission over large enough distances,
 e.g., the end-to-end transmission, see \cref{fig:conceptual_figure}(a). 
 
\begin{figure}
    \centering
    \includegraphics[width=\linewidth]{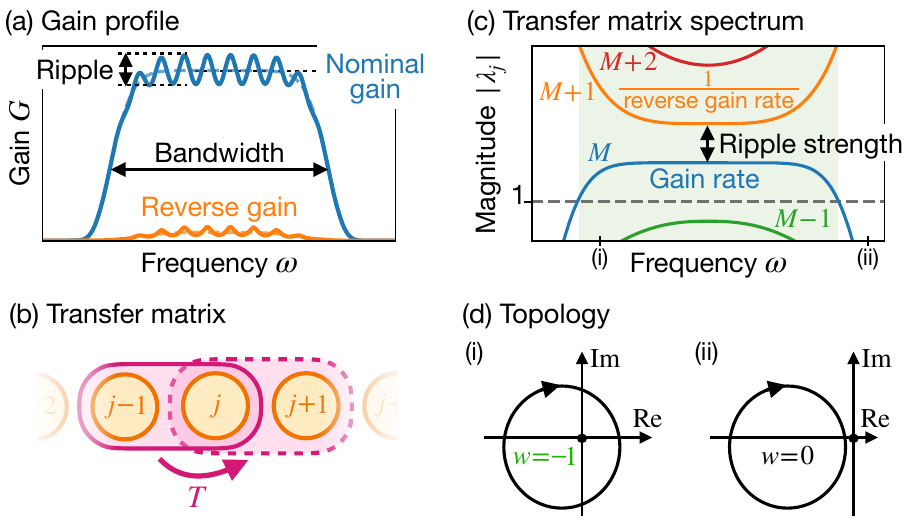}
    \caption{
    \textbf{Connection between the gain properties and the transfer matrix.}
    (a) Gain factor (blue) and reverse gain (orange) as a function of the input frequency.
    (b) Transfer matrix approach. The transfer matrix describes how excitations on the current and previous unit cell are passed on to the next unit cell.
    (c) Spectrum of the transfer matrix. The $M$th smallest eigenvalue is the gain rate per unit cell, the $(M+1)$th smallest eigenvalue is the inverse of the reverse gain rate. 
    The distance between the $M$th smallest eigenvalue and its nearest neighbors determines the ripple.
    (d) Connection between the transfer matrix and non-Hermitian topology, shown for two exemplary points from (c). The curve shows the real and imaginary part of $\det(H(k)-\omega\mathds{1})$ as a function of $k$. In (i), $\abs{\lambda_M}$ is greater than \num{1} (green-shaded area in (c)), resulting in a topological winding number $w=-1$, see \cref{eq:winding}. In (ii), $\abs{\lambda_M}$ and $1/\abs{\lambda_{M+1}}$ are smaller than \num{1} resulting in a winding number $w=0$.
    }
    \label{fig:gain_profile}
\end{figure}

For concreteness, consider a directional amplifier as an example. 
The typical  gain, $G(\omega)=|S_{\mathbf{j}_{\rm out}\mathbf{j}_{\rm in}}(\omega)|^2$, and reverse gain,  $|S_{\mathbf{j}_{\rm in}\mathbf{j}_{\rm out}}(\omega)|^2$, are shown in \cref{fig:gain_profile}(a). Key features that characterize the amplifier include the nominal gain, the amplification bandwidth, the amplitude of the ripples modulating the gain within the amplifier bandwidth, and the maximum reverse gain. A naive approach would optimize these key quantities directly for a fixed system size. However, this would lead to results that are not transferable to different system sizes, as most of these quantities actually depend on $N$. Instead, our approach exploits the fact that all key quantities have a well-defined scaling with the number of unit cells and directly optimizes this scaling. For example, the nominal gain increases exponentially with $N$, which allows us to define the gain rate per unit cell $s(\omega)=\lim_{N\to \infty} (\ln |S_{\mathbf{j}_{\rm out}\mathbf{j}_{\rm in}}(\omega)|/N)$. We can define similar rates for the reverse gain and the ripple strength (as they decrease exponentially with size). In other examples, we adapt the key features to the target functionality, but always use the same type of rates. 
In all cases, we want the gain rate per unit cell $s$ to adopt a certain value $s_\mathrm{target}$ at a target frequency $\omega_\mathrm{target}$. Furthermore, to optimize the bandwidth, we prefer the frequency dependence to be flat, which we can ensure by requiring that the derivatives of the gain rates be zero up to a certain order. Similarly, we can enforce conditions for other asymptotic gain properties. In addition to these types of constraints, we also require that all lattices are dynamically stable.

Each of those constraints can be expressed by a function $f(\vec{x})$, where $\vec{x}$ denotes all free continuous parameters of a chosen lattice model, like its coupling rates $g$, and $\nu$, and loss rates $\kappa$. For example, we use the function $f(\vec{x})=s(\vec{x})-s_\mathrm{target}$ to enforce a target gain rate at some desired frequency $\omega$. We choose the loss function
\begin{equation}
    \label{equ:loss_function}
    \mathcal{L} = \sum_j \abs{f_j}^2,
\end{equation}
which sums up all selected target features. For an overview of all target features and their implementation, see \cref{app:optimization_constraints}.



Given a certain lattice model, the continuous optimization minimizes the loss function in \cref{equ:loss_function} with respect to the free parameters $\vec{x}$. We are looking for lattices that fulfill the desired characteristics perfectly, so there exists a certain parameter set $\vec{x}^\ast$ for which $\mathcal{L}(\vec{x}^\ast)=0$. We label all those lattices as valid lattices. For more details on the optimization, we refer to \cref{app:details_continuous_optimization}.\\




\subsection*{Transfer-matrix approach. }  
We now introduce our transfer-matrix approach, which provides access to the scaling of the gain properties in the thermodynamic limit. 
To anticipate our most important result: We show that the gain rate per unit cell equals the $M$th smallest eigenvalue of the transfer matrix, and that the reverse gain equals the inverse of the $(M+1)$th smallest eigenvalue, see \cref{fig:gain_profile}(c). This result is closely linked to other concepts from non-Hermitian topology, particularly the scaling law presented in \cite{xue2021simple}, which we discuss below and in more detail in \cref{app:connection_to_topology}.


Transfer matrices are a well-established tool for characterizing Hermitian one-dimensional periodic systems. They are commonly employed to analyze, inter alia, the transport properties \cite{yeh1977electromagnetic}, band structures \cite{lee1981simple,chang1982complex,hatsugai1993edge}, and topology \cite{hatsugai1993edge,wielian2025transfer} of these systems. 
In the context of non-Hermitian systems, transfer-matrix approaches are less established. Nevertheless, important connections have recently been established between transfer-matrix methods and key concepts in non-Hermitian physics \cite{kunst2019non, koekenbier2024transfer}. 
In the following, we show the connection between the transfer matrix and the scattering matrix of non-Hermitian systems under open boundary conditions (OBC), and how the spectrum of the transfer matrix determines the scaling of the gain properties in the thermodynamic limit.

For simplicity, we focus on local couplings, up to the neighboring unit cells. Furthermore, we apply our method to phase-preserving devices. This restricts the available couplings; for example, on-site squeezing is not allowed, see \cref{app:full_Hamiltonian} for more details. For this class of systems, we can write the Langevin equations in the compact form
\begin{equation}
    \label{equ:equations_of_motion}
    \dot{\vec{a}}_j = \mu_{-1}\vec{a}_{j-1} +\mu_{0}\vec{a}_{j} + \mu_{1}\vec{a}_{j+1}-\sqrt{\kappa} \vec{a}_{j,\mathrm{in}}
\end{equation}
for $j=1,\ldots,N$. Here, the operator vectors $\vec{a}_j$ have $M$ entries corresponding to an annihilation or a creation operator on each sublattice, e.g., $\vec{a}_j=(\hat{a}_{j,1},\hat{a}^\dagger_{j,2})$. To extend the validity of the equations to the boundaries, we have also defined $\vec{a}_{0}=\vec{a}_{N+1}=0$. Moreover, we have grouped all decay rates, detunings, and couplings in the matrices $\kappa=\text{diag}(\kappa_1, \ldots, \kappa_M)$, and  $\mu_{l}$, $l=-1,0,1$, see Methods for the exact expressions. 

In the following, we assume for simplicity that $\mu_{\pm 1}$ are invertible matrices. In \cref{app:invertiblity_problem}, we demonstrate that our core statements still hold when $\mu_{\pm 1}$ is not invertible. 

After going into frequency space and solving \cref{equ:equations_of_motion} for $\vec{a}_{j+1}$, we can rewrite the equations of motion as
\begin{equation}
    \label{equ:equations_of_motion_T}
    \begin{pmatrix}\vec{a}_{j+1} \\ \vec{a}_j  \end{pmatrix} = T
\begin{pmatrix}\vec{a}_{j} \\ \vec{a}_{j-1}  \end{pmatrix} + \begin{pmatrix}\mu_1^{-1} \sqrt{\kappa} \vec{a}_{j,\mathrm{in}} \\ 0\end{pmatrix},
\end{equation}
with the transfer matrix $T$ defined as
\begin{equation}
    \label{equ:define_transfer_matrix}
    T(\omega) = 
    \begin{pmatrix}
        - \mu_1^{-1} (\mu_0 + i \omega \mathds{1}) & -\mu_1^{-1}\mu_{-1}\\
        \mathds{1} & 0
    \end{pmatrix}.
\end{equation}
Here, $\mathds{1}$ and $0$ denote the $M\times M$ dimensional identity and zero matrix and $\omega$ the frequency in Fourier space. The transfer matrix $T$ describes how excitations on the current and previous unit cells $j$ and $j-1$ are passed on to the next unit cell $j+1$, see \cref{fig:gain_profile}(b). \\

\subsection*{Relation to the scattering matrix. } \label{sec:input_output_relation} In this section, we use the transfer matrix to calculate the left-end-to-right-end scattering matrix $S_\mathrm{RL}(\omega)$. The scattering matrix element $S_{\mathrm{RL},ji}(\omega)$ describes the transmission of an input signal injected at mode $i$ on the first unit cell to the output extracted at mode $j$ on the last unit cell. 
For simplicity, we assume that there is only an input signal at the first unit cell, and that all other inputs are zero. In Methods, we demonstrate how to calculate any other scattering paths, also from and to intermediate unit cells.



For $j=1$, we rewrite \cref{equ:equations_of_motion_T} as:
\begin{equation}
    \label{equ:derive_forward_scattering_j_1}
    \begin{pmatrix}\vec{a}_{1} \\ 0  \end{pmatrix} = T^{-1} \begin{pmatrix} \vec{a}_{2} \\ \vec{a}_{1}  \end{pmatrix} + \begin{pmatrix} 0 \\ \mu_{-1}^{-1} \sqrt{\kappa} \vec{a}_{1,\mathrm{in}} \end{pmatrix}.
\end{equation}
Note that $\vec{a}_{0}=0$ to fulfill the OBC. We now iteratively insert \cref{equ:equations_of_motion_T} until we arrive at the rightmost unit cell $j=N$. Here, we obtain
\begin{equation}
    \label{equ:to_solve_for_gain}
    \begin{pmatrix} \vec{a}_1 \\ 0  \end{pmatrix} = 
T^{-N}
\begin{pmatrix}0 \\ \vec{a}_{N} \end{pmatrix} +\begin{pmatrix} 0\\\mu_{-1}^{-1} \sqrt{\kappa} \vec{a}_{1,\mathrm{in}} \end{pmatrix},
\end{equation}
also considering that $\vec{a}_{N+1}$ equals $0$ due to the OBC.
Solving for $\vec{a}_N$ and using the boundary condition 
$\vec{a}_{j,\mathrm{out}} = \vec{a}_{j,\mathrm{in}} + \sqrt{\kappa} \vec{a}_j$, we calculate the output field $\vec{a}_{N,\mathrm{out}}$. Therefore, we can express the end-to-end the scattering matrix as
\begin{equation}
    \label{equ:scattering_left_to_right}
    S_\mathrm{RL}(\omega) = -\sqrt{\kappa}(PT^{-N}P^t)^{-1}\mu_{-1}^{-1}\sqrt{\kappa}.
\end{equation}
Here, $P=(0, \mathds{1})$ is a $M\times2M$ matrix, and $P \bullet P^t$ selects the lower right block of a matrix $\bullet$. $t$ denotes the transpose.

To extract the scaling of the gain properties from \cref{equ:scattering_left_to_right}, we diagonalize the transfer matrix, where $\lambda_{m=1,\ldots,2M}$ denote its eigenvalues ordered according to their magnitude $\abs{\lambda_1} \leq \abs{\lambda_2} \leq \ldots \leq \abs{\lambda_{2M}}$. 
We find that, to leading order, the scattering matrix scales as:
\begin{equation}
    \label{equ:scattering_approximate}
    S_\mathrm{RL}(\omega) = A(\omega) \lambda_M (\omega) ^ N.
\end{equation}
The basis of this exponential scaling is the $M$th smallest eigenvalue $\lambda_M$ which can be interpreted as a gain rate per unit cell, see \cref{fig:gain_profile}(c). Note that the larger eigenvalues $\lambda_{m>M}$ do not describe the amplification of waves going to the right but rather the inverse gain of waves moving leftwards. 
Consequently, repeating the same analysis for the reverse direction shows that the reverse gain scales with $1/\lambda_{M+1}^N$. Therefore, we define $1/\lambda_{M+1}$ as the reverse gain rate. We also find the explicit expression for the scaling prefactor $A(\omega)$, which depends on the eigenvectors of the transfer matrix, see \cref{app:derivation_gain_scaling} for details.

Higher-order corrections to \cref{equ:scattering_approximate} give rise to the ripples. The ripples are exponentially suppressed in the long-lattice limit. Their decay rate is determined by $\abs{\lambda_{M-1}/\lambda_{M}}$, or $\abs{\lambda_{M}/\lambda_{M+1}}$, depending on which of these ratios is larger; see \cref{app:derivation_gain_scaling} for more details. \\




\subsection*{Connection to non-Hermitian topology. } 

It was recently shown~\cite{Porras2018,wanjura2020topological} that directional amplification, in particular the exponential scaling of the end-to-end gain, is in one-to-one correspondence with a non-trivial notion of non-Hermitian topology. 
In this section, we show that the transfer matrix has a deep connection to non-Hermitian topology and allows us to unify multiple different approaches for treating non-Hermitian topological systems.

We start reviewing the notion of the topological invariant for a non-Hermitian $1D$ Bloch Hamiltonian $H(k)$ with a point gap~\cite{gong2018topological,kawabata2019}. As usual, $H(k)$ is derived assuming periodic boundary conditions (PBC) and is parametrized by the quasimomentum $k$.
As the quasi-momentum $k$ varies over the Brillouin Zone (BZ), the determinant  $\det(H(k)-\omega\mathds{1})$ with $\omega$ inside a point gap (i.e., $\omega$ is not an eigenvalue of $H(k)$ for any $k$) traces a closed loop in the complex plane without crossing the origin. Thus, the winding number
\begin{equation}
    \label{equ:winding_number}
    w = \frac{1}{2\pi i} \int_{-\pi}^\pi \mathrm{d}k \frac{\partial}{\partial k} \ln \det(H(k)-\omega\mathds{1}),
\end{equation}
counting how many times the loop encircles the origin, is a topological invariant associated to the point gap.

In our setting of dissipative lattice models, the non-Hermitian Hamiltonian $H(k)$ of interest is derived from the homogeneous part of Eq.~(\ref{equ:equations_of_motion}) when extending its validity to arbitrary integer $j$, i.e., we consider the deterministic dynamics of an infinite chain without input fields. By inserting the Bloch wave ansatz
\begin{equation}
    \label{equ:Bloch_ansatz}
\vec a_{j}(t)=e^{i (k j-E t)}\vec a_k
\end{equation}
into the homogeneous part of Eq.~(\ref{equ:equations_of_motion}), we find the non-Hermitian eigenvalue problem
\begin{equation}
    \label{equ:Bloch_ham}
H(k)\vec a_k=E{\vec a}_k,\quad H(k) = i\mu_{-1} e^{-\mathrm{i}k} + i\mu_0 + i\mu_1 e^{\mathrm{i}k}.
\end{equation}

As discussed in Ref.~\cite{wanjura2020topological}, for $\omega$ on the real axis, the sign of the winding number in \cref{equ:winding_number} determines the direction of the amplification of a signal at the corresponding frequency: For negative $w$, signals transmitted from left to right are amplified. For positive $w$, the amplification direction is reversed. For trivial topology $w=0$, signals are de-amplified in both directions.

Next, we analyze the connection of the transfer matrix method to the non-Hermitian Hamiltonian $H(k)$ and its winding number $w$. We note that by recursively applying the transfer matrix $T(\omega)$ to one of its eigenvectors with eigenvalue $\lambda_m$ one can construct a solution of the homogeneous part of Eq.~(\ref{equ:equations_of_motion}) that also fulfills the ansatz \cref{equ:Bloch_ansatz} with energy $E=\omega$ and complex quasimomentum  $k=-i \ln \lambda_m$, see \cref{equ:equations_of_motion_T} with ${\vec a}_{j,\rm in}=0$. Since no specific boundary condition is imposed, one can choose an arbitrary $\omega$ without any constraint from the PBC point gap and always find $2M$ solutions with their $\lambda_m$. These solutions also correspond to right eigenvectors of the generalized Bloch-Hamiltonian $H(k)$ with complex $k$, $k=-i\ln\lambda_m$. They are formed by the first $M$ entries of the corresponding $T$-matrix eigenvectors, i.e., the amplitudes within one unit cell. 
Most importantly,  the winding number $w$ is directly accessible from the transfer matrix formalism via the number of $\lambda_m$ within the unit circle  (see \cref{app:connection_to_topology}):
\begin{equation}
    \label{eq:winding}
    w=(\text{number of $\abs{\lambda_m} < 1$}) - M.
\end{equation}

The transfer matrix approach is naturally connected to the generalized Brillouin zone (GBZ) approach~\cite{Yao2018edge,yokomizo2019non,yang2020non}. Also in this approach, one views the non-Hermitian Hamiltonian $H(k)$ as a function of an arbitrary complex number $\lambda=e^{ik}$. In this case, one focuses on the complex spectrum $\omega$ for open boundary conditions (OBC). The central result is that for $\omega$ in the OBC spectrum, after we order the $\lambda_j$ by their absolute value, we find  $\lvert\lambda_M(\omega)\rvert=\lvert\lambda_{M+1}(\omega)\rvert$.
This connection has important implications for the onset of a dynamic instability. Crossing points between different eigenvalues $\lvert\lambda_M(\omega)\rvert$ and  $\lvert\lambda_{M+1}(\omega)\rvert$ at some real value $\omega$ mean that at this point $\lambda_M(\omega)$ lies on the GBZ, so the frequency $\omega$ must be an OBC eigenvalue with zero imaginary part. This implies that this point is the onset of a dynamic instability.
We use this in our optimisation to ensure we only discover stable lattice models. 

In the following, we discuss several illustrative examples of lattices designed with our approach that fulfill different wave transport functionalities. 
The models we discover for the directional amplifiers and frequency demultiplexers also have an underlying non-trivial topology. Therefore, our scheme also enables the discovery of systems with certain non-Hermitian topological properties. \\

\begin{figure}
    \centering
    \includegraphics[width=\linewidth]{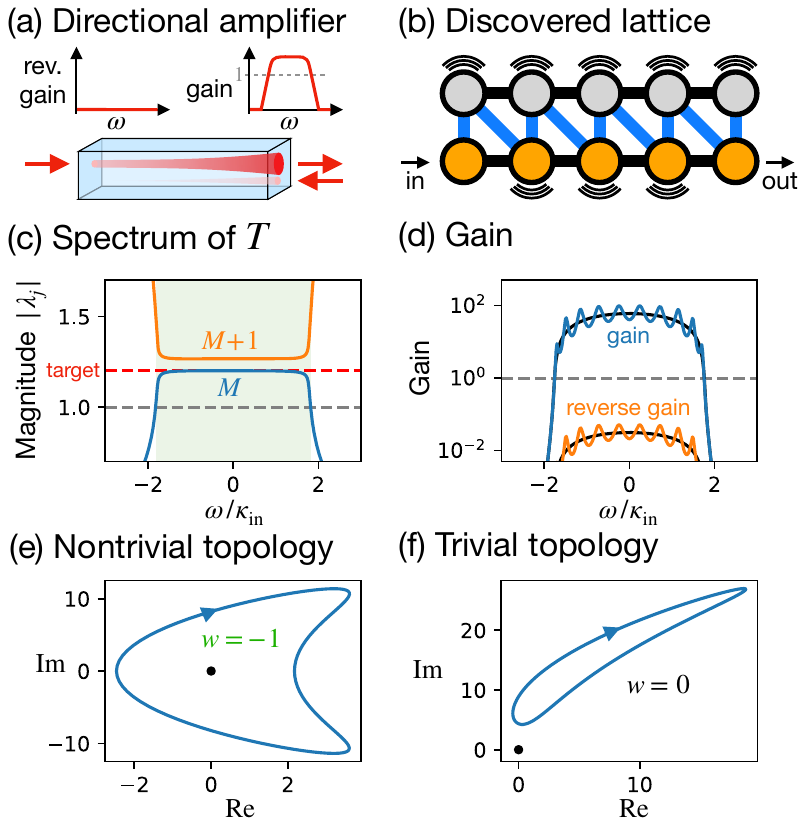}
    \caption{\textbf{Designing an amplifier with constant gain rate over a wide bandwidth.} (a) Target behavior. The device is designed to amplify a signal from left to right, while attenuating in the reverse direction. The forward gain should remain as constant as possible across the frequency range of interest. (b) One of the discovered lattices, see Supplemental Material for an exhaustive list. (c) Transfer matrix spectrum of the lattice in (b). The green-shaded area is the frequency range in which $\abs{\lambda_M}$ is greater than \num{1}. (d) Forward and reverse gain profiles (blue and orange) for the lattice in (b) with \num{10} unit cells. (e,f) Topology of the designed lattice. In (e), $\omega$ equals $0$ (within the green shaded area in (c)). In (f), $\omega$ equals $2.5\kappa_\mathrm{in}$ (outside of the green-shaded area).}
    \label{fig:amplifier}
\end{figure}

\subsection*{Designing an amplifier with constant gain rate over a wide bandwidth. } As a first example, we consider a directional amplifier that provides broadband amplification in the thermodynamic limit, see \cref{fig:amplifier}(a). There exists a multitude of proposed schemes that only rely on a handful of modes to achieve directional and phase-preserving amplification, see \cite{metelmann_nonreciprocal_2015,ranzani_graph-based_2015,lecocq_nonreciprocal_2017,malz_quantum-limited_2018,sliwa2015reconfigurable,abdo2013directional,liu_fully_2023,AutoScatter}. However, the bandwidth of all of those devices decays with the gain $G$, e.g., the amplifier's bandwidth in \cite{metelmann_nonreciprocal_2015} decays with $1/\sqrt{G}$. 
It is known that periodic structures can overcome this challenge and achieve much better gain-bandwidth products \cite{ho2012wideband}. However, ultimately, the usable bandwidth of the gain profile is determined by the flatness of the gain rate, as any deviation will be amplified exponentially, see \cref{equ:scattering_approximate}.

Therefore, to design a high-bandwidth amplifier, we impose the condition that several derivatives of the gain rate with respect to $\omega$ vanish.
Specifically, we set the first three derivatives to zero, balancing the goal of attaining a sufficiently flat gain rate profile with the need for numerical stability in computing these derivatives, see \cref{app:optimization_constraints}.

The effect of the scaling prefactor $A$ is weaker. However, we find that the quality of the discovered amplifiers is much better when we also enforce that the first derivative of $A$ with respect to $\omega$ vanishes. Furthermore, we enforce that the lattices' dynamics are stable (i.e., the imaginary part of all eigenvalues of $H$ is negative), and that there is a minimum distance between the magnitudes of $\lambda_M$ and $\lambda_{M + 1}$, reducing the ripple strength, see \cref{fig:gain_profile}(a,c). 


Finally, we enforce amplification by imposing that the gain rate per unit cell at $\omega=0$ reaches a specified target value $s_\mathrm{target}$, with $s_\mathrm{target}>1$. After discovering suitable lattice models, we vary the value of $s_\mathrm{target}$, and use symbolic regression to analyze the dependencies of the lattices' parameters on this target value, see the next section.

Our approach discovers multiple lattice models fulfilling those constraints; see Supplemental Material for an exhaustive list of all discovered lattices. Of those, we discuss the lattice in \cref{fig:amplifier}(b) in more detail as its gain rate profile is exceptionally flat, see \cref{fig:amplifier}(c). Its first five derivatives are smaller than \num{1e-2}, which exceeds the constraints we set and leads to broadband amplification; see \cref{fig:amplifier}(d). For comparison, for a Josephson traveling wave amplifier, only the first derivative of the gain rate vanishes \cite{white2015traveling}.




The discovered amplifier also demonstrates the connection between the eigenvalues of the transfer matrix and topology. When $\abs{\lambda_M}$ is larger than \num{1}, the winding number is $w=-1$ and the lattice amplifies signals towards the right, see \cref{fig:amplifier}(e). When $\abs{\lambda_M}$ and $1/\abs{\lambda_{M+1}}$ are both smaller than \num{1}, the lattice is in a topologically trivial state with $w=0$, and signals are deamplified in both directions, see \cref{fig:amplifier}(f). \\




\subsection*{Symbolic Regression. }
\label{sec:symbolic_regression}

Symbolic expressions describe the functional dependencies between physical quantities and naturally provide interpretability, which aids applications and generalization. Therefore, human researchers often seek analytical solutions or approximations for a physical system. 
When mathematical models are not available or too complex, symbolic expressions are typically derived through heuristic methods, e.g., by fitting various candidate functions. Symbolic regression \cite{schmidt2009distilling,brunton2016discovering,udrescu2020ai} provides a powerful, automated approach to this problem. Unlike traditional regression, which fits continuous parameters to a fixed mathematical expression, symbolic regression searches through both the space of possible expressions and their associated parameters. In any symbolic regression task, there is a trade-off between the complexity of an expression and its fit error. We aim to discover the set of all Pareto-optimal expressions, in the sense of finding the expressions with the smallest fit error as a function of the complexity.

In our work, we use {\sc AIFeynman 2} \cite{AIFeynman2,github_aifeynman}, which is specifically tailored to physics applications and discovers and exploits symmetries in the dataset. Using {\sc AIFeynman 2}, we 
automatically identify Pareto-optimal symbolic relationships between the lattice parameters and idealized gain properties in our discovered amplifier lattices. These equations allow us to gain insights into the underlying parameter dependencies and to engineer amplifiers with target gain characteristics without re-running the optimization. This is especially helpful for lattices where we were unable to obtain analytical solutions ourselves, e.g., for the lattice in \cref{fig:amplifier}(b), which we use as a primary example below.

\begin{figure}
    \centering
    \includegraphics[width=\linewidth]{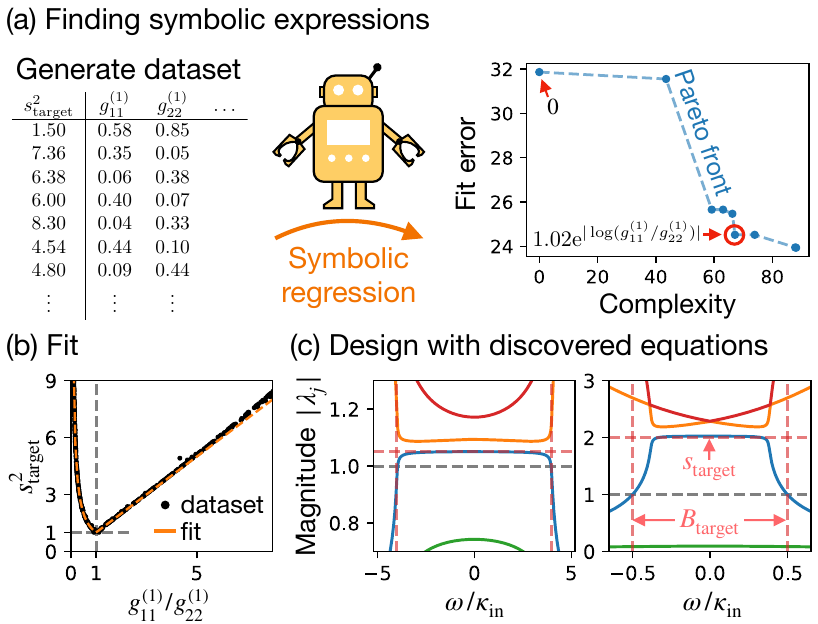}
    \caption{\textbf{Automated discovery of symbolic expressions of the optimized lattice models.} (a) Pathway to find symbolic expressions. 
    Given a candidate lattice model (here \cref{fig:amplifier}(b)), we run the continuous optimization for different target gain rates $s_\mathrm{target}$, creating a dataset of gain rates and lattice model parameters. Using the symbolic regression tool \textsc{AIFeynman 2}, we find symbolic expressions that are Pareto-optimal, i.e., having the smallest fit error given a certain complexity. The fit error is defined as the mean error description length, and complexity as the description length of the symbolic expression, see \cite{AIFeynman2,wu2019toward}.
    (b) Comparison between the dataset values and \cref{equ:gain_rate_equation} found via symbolic regression.
    (c) Design of amplifiers with a target gain rate per unit cell and target bandwidth (dashed red lines) using the discovered symbolic expressions in \cref{equ:gain_rate_equation,equ:bandwidth_equation,equ:sqeezing_rates_1,equ:sqeezing_rates_2}. We engineer an amplifier with a large bandwidth and small gain rate per unit cell (left), and with a large gain rate and a small bandwidth (right).}
    \label{fig:symbolic_regression}
\end{figure}

To utilize symbolic regression, we have to generate a large dataset covering a wide range of target gain characteristics and the associated lattice parameters, see \cref{fig:symbolic_regression}(a). To do so, we run our continuous optimization multiple times for different target gain rates while keeping the lattice layout fixed.

Simultaneously, we extract for every parameter set the bandwidth, a quantity which we have so far controlled only indirectly by constraining the derivatives of the gain rate per unit cell. We define the bandwidth of our amplifiers as the frequency range in which amplification occurs, meaning that the gain rate per unit cell $\abs{\lambda_M}$ is greater than \num{1}, or equivalently, the winding $w$ is negative (green-shaded area in \cref{fig:amplifier}(c)). In contrast to other common definitions, like the full-width-at-half-maximum, our definition offers the advantage of being independent of the lattice size (for sufficiently large lattices).

We use {\sc AIFeynman 2} \cite{AIFeynman2,github_aifeynman} to find Pareto-optimal expressions for the bandwidth $B$ and the gain rate squared $s^2$ of the lattice in \cref{fig:amplifier}(b). They depend on the magnitude of the remaining coupling parameters $g_{11}^{(1)}$, $g_{22}^{(1)}$, $\nu_{12}^{(0)}$, and $\nu_{21}^{(1)}$, see Methods for their definition. For simplicity, we focus our analysis on the case where the losses of the modes in a unit cell are equal to each other, so $\kappa_1=\kappa_2$, thereby reducing the number of free variables. We chose to search for the gain rate squared $s^2$ instead of the gain rate $s$, as $s^2$ describes directly the scaling of the gain $G$. 


For the gain rate squared, we find a multitude of expressions, ranging from the simplest function $s_\mathrm{target}^2=0$ up to $s_\mathrm{target}^2=1.004 + \Big(\frac{g_{11}^{(1)} + g_{22} ^{(1)}}{2 \nu_{12}^{(0)} + 2 \nu_{21}^{(1)} - 1} - \frac{1}{\pi - \log{\pi }}\Big)^{-1}$, see \cref{fig:symbolic_regression}(a). Especially, expressions at sudden jumps of the Pareto front plot are useful as they offer a significant improvement in accuracy for only a small increase in complexity, e.g., the discovered expression $s_\mathrm{target}^2 = 1.02 \mathrm{exp} ( | \mathrm{log} ( g_{11}^{(1)} / g_{22}^{(1)} ) | )$. See \cref{fig:symbolic_regression}(b) for a comparison with the dataset. We can rewrite this expression more compactly in the form:
\begin{equation}
    \label{equ:gain_rate_equation}
    s_\mathrm{target}^2 = \max \left(\frac{g_{11}^{(1)}}{g_{22}^{(1)}}, \frac{g_{22}^{(1)}}{g_{11}^{(1)}}\right),
\end{equation}
where we also rounded the prefactor to \num{1}.
{\sc AIFeynman 2} was not able to discover this form as the $\max$ function is not available as a building block. Interestingly, this expression is independent of the squeezing rates. This means that the ratio of the beamsplitter coupling rates primarily determines the gain rate.

Repeating the same analysis for the bandwidth, we find after a sudden jump the expression
\begin{equation}
    \label{equ:bandwidth_equation}
    B_\mathrm{target} = 2 g_{11}^{(1)} + 2g_{22}^{(1)},
\end{equation}
which again only depends on the beamsplitter coupling rates.

To obtain the inverse design of an amplifier given a specific target gain rate $s_\mathrm{target}^2$ and target bandwidth $B_\mathrm{target}$, we can now solve \cref{equ:gain_rate_equation,equ:bandwidth_equation} for $g_{11}^{(1)}$ and $g_{22}^{(1)}$. Reapplying {\sc AIFeynman 2} to find expressions for the missing squeezing rates, we find in good approximation:
\begin{gather}
    \label{equ:sqeezing_rates_1}
    \nu_{12}^{(0)} = 0.421,  \\
    \label{equ:sqeezing_rates_2}
    \nu_{21}^{(1)} = 0.5 - \nu_{12}^{(0)} +\abs{ g_{11}^{(1)} - g_{22}^{(1)}}.
\end{gather}
Lastly, we find that the synthetic field fluxes enclosed in any triangular loop between nearest neighbors equal $\pi/2$.

Now, we have all the ingredients required for the inverse design of amplifiers. As a demonstration, we use \cref{equ:gain_rate_equation,equ:bandwidth_equation,equ:sqeezing_rates_1,equ:sqeezing_rates_2} to engineer an amplifier with a small gain rate and large bandwidth, and another one with a large gain rate and small bandwidth, see \cref{fig:symbolic_regression}(c). \\











\subsection*{Designing an isolator. }

\begin{figure}
    \centering
    \includegraphics[width=\linewidth]{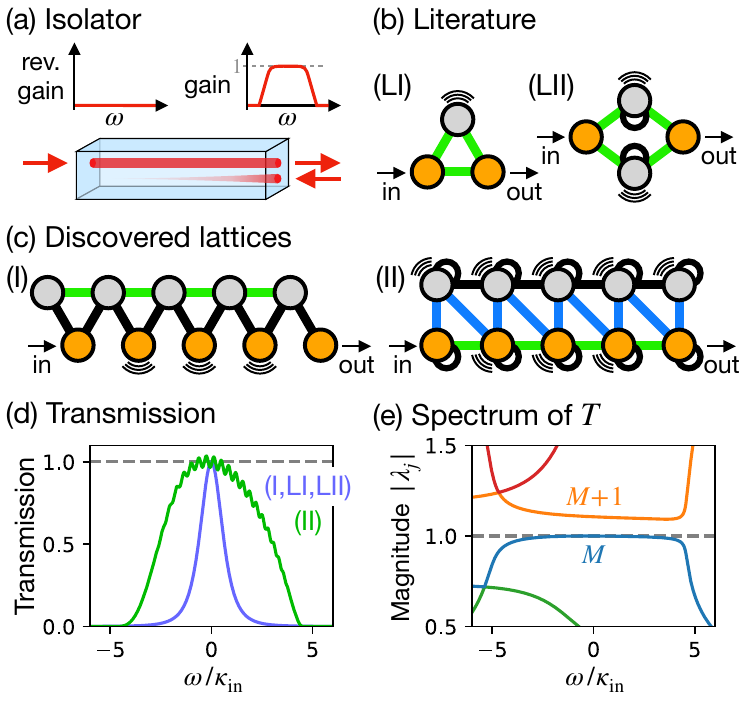}
    \caption{
    \textbf{Automated design of an isolator.}
    (a) Target behavior. The device transmits a signal from left to right with unity transmission, while suppressing the reverse direction.
    (b) Previously proposed few-mode isolator schemes. (LI), proposed in \cite{lecocq_nonreciprocal_2017,sliwa2015reconfigurable,habraken_continuous_2012}, couples input and output (orange) to one auxiliary mode (gray). In the following, we consider the case where the auxiliary mode is overdamped and acts as a bath, see \cite{metelmann_nonreciprocal_2015}.
    In (LII), the input and output are microwave modes (orange) coupled via two strongly detuned mechanical modes (gray), see \cite{bernier2017nonreciprocal}.
    (c) Two of the discovered lattice models, see Supplemental Material for an exhaustive list. (I) only uses beamsplitter couplings, (II) also incorporates squeezing. The gray modes in (I) are lossless.
    (d) Transmission profile of the isolator schemes from (b) and (c). 
    To ensure a fair comparison, the in- and out-coupling rates of all schemes are set to $\kappa_\mathrm{in}$. The schemes (LI), (LII), and (I) share the same transmission spectrum (blue curve). The reverse transmission (not shown) vanishes for (LI) and (LII). For the lattices (I) and (II), it is exponentially suppressed in the thermodynamic limit. The length of these lattices is set to $N=40$.
    (e) Spectrum of the transfer matrix for lattice (II).
    }
    \label{fig:isolator}
\end{figure}

Another important task in signal processing is isolation, i.e., a signal is perfectly transmitted in one direction, while the other direction is blocked. This type of device helps suppress unwanted noise that leaks back to the signal source.

To design novel isolator schemes, we express this task in terms of the idealized gain properties. For perfect transmission, both the gain rate and scaling prefactor have to be \num{1}. Furthermore, we want many of their derivatives to be zero, especially those of the gain rate. 
To ensure small ripples, we enforce that $\abs{\lambda_{M+1}}$ keeps a minimum distance from $\abs{\lambda_{M}}$. This automatically ensures reverse isolation, as it enforces $\abs{\lambda_{M+1}}>1$, leading to a reverse gain rate smaller than \num{1}. Lastly, we demand that the discovered lattices are dynamically stable.

When we run our optimization for this scenario, we discover two distinct types of isolators, see \cref{fig:isolator}(c). The first group (I) uses only beamsplitter couplings, while the second group (II) incorporates both beamsplitter and squeezing interactions. The two lattices shown are characteristic examples of those discovered. See the Supplemental Material for a complete list of all the lattices.

In \cref{fig:isolator}(d) we compare both lattices to previously proposed isolators based on a few coupled modes \cite{habraken_continuous_2012,sliwa2015reconfigurable,lecocq_nonreciprocal_2017,bernier2017nonreciprocal,metelmann_nonreciprocal_2015}, see \cref{fig:isolator}(b). 
We find that our discovered lattice (I) shares the same transmission profile as the previous proposals (LI) and (LII), and does not offer any advantage compared to them. In contrast, (LII) has a better bandwidth than any of the other schemes.


Lastly, we use {\sc AIFeynman 2} to automatically discover the underlying parameter dependencies for lattice (I). As the target gain rate per unit cell is fixed to \num{1}, and the bandwidth is almost constant, we only have to fit the dependencies between the remaining system parameters $g_{12}^{(0)}$, $g_{12}^{(1)}$, $g_{22}^{(1)}$, and $\kappa_2$. This system is simpler than the amplifier above, and we were able to derive an exact analytical solution ourselves. {\sc AIFeynman} either successfully rediscovers our expressions or equivalent transformations of them, which constitutes an important check of the automated approach. The coupling rates connected to the orange mode in \cref{fig:isolator}(c) have to be  equal, $g_{12}^{(0)}=g_{12}^{(1)}$, and the remaining coupling rate $g_{22}^{(1)}$ has to be $g_{22}^{(1)} = 2 (g_{12}^{(0)})^2$. The loss rate of the gray modes $\kappa_2$ is zero. Lastly, the synthetic field fluxes enclosed in the triangular loops are $\pi/2$. \\




\subsection*{Designing a frequency demultiplexer. }

\begin{figure}
    \centering
    \includegraphics[width=\linewidth]{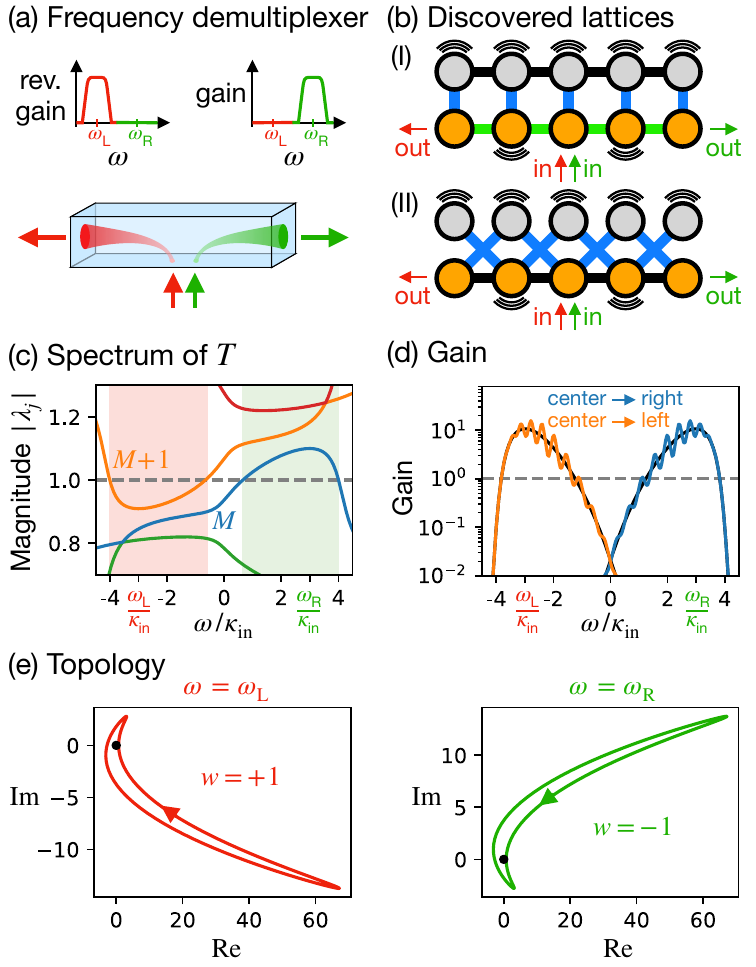}
    \caption{\textbf{Automated design of a frequency demultiplexer.} (a) Target behavior. The device is designed to split up two incoming signals at the frequencies $\omega_\mathrm{L}$ (red) and $\omega_\mathrm{R}$ (green), making them travel in different directions.
    (b) Irreducible lattice models discovered by our approach. (c) Transfer matrix spectrum of lattice (I) in (b). (d) Corresponding gain profiles for scattering from the central unit cell to the left (orange) and right (blue) with \num{51} unit cells. The idealized gain profiles are shown in black. (e) Topological properties of the designed lattice. The winding number equals $\pm 1$ red/green-shaded area in (c).}
    \label{fig:frequency_splitter}
\end{figure}

Lastly, we show an example where the frequency dependence is more complex than before: we use our approach to design a frequency demultiplexer. The goal is to divide an input signal into two distinct frequency components, each forwarded to opposite ends of the lattice, see \cref{fig:frequency_splitter}(a). For symmetry, the input signal enters at the center of the lattice.

To design such a system, we demand that the forward gain rate $\abs{\lambda_M}$ equals some target value $s_\mathrm{target}>1$ at the frequency $\omega_\mathrm{R}$, which is the center of the frequency band intended for transmission to the right end. To achieve symmetric amplification, the reverse gain rate $1/\abs{\lambda_{M+1}}$ should equal the same value $s_\mathrm{target}$ at the center frequency $\omega_\mathrm{L}$ of the band sent to the left. 
Additionally, we impose that the first derivatives of both gain rates and associated scaling prefactors vanish at their respective target frequencies, ensuring flat gain profiles. 
We note that the definition of the scaling prefactor changes compared to the previous examples, as we consider here the scattering from an intermediate unit cell to one of the boundaries. The definitions of the scaling rates are not affected by this; see \cref{app:deriviation_scaling_intermediate} for more details. 
Lastly, we enforce a minimum separation between the central eigenvalues and the lattice to ensure dynamic stability.

With these constraints, we discover the two lattice models shown in \cref{fig:frequency_splitter}(b). The eigenvalue spectrum of the transfer matrix exhibits the desired behavior, see \cref{fig:frequency_splitter}(c), leading to a separation of the input frequencies, see \cref{fig:frequency_splitter}(d). 
The frequency-demultiplexer behavior has important implications for the lattice's topology. In the frequency range where the gain rate $\abs{\lambda_{M}}$ is larger than \num{1} (green-shaded area in \cref{fig:frequency_splitter}(c)), the winding number is $w=-1$ (amplification to the right). In the area where the reverse gain rate $1/\abs{\lambda_{M+1}}$ is larger than \num{1} the winding is $w=1$ (amplification to the left), see \cref{fig:frequency_splitter}(e). Everywhere else, the winding number is zero. \\





\begingroup
\let\MakeTextUppercase\relax
\section*{Discussion}
\endgroup
\noindent In this work, we have developed an approach to automatically discover one-dimensional lattice models with desired wave transport functionalities. Our method optimizes the discrete and continuous lattice parameters and provides an exhaustive list of all possible setups fulfilling the target characteristics, up to a certain complexity. At its core, it leverages a transfer-matrix approach that characterizes the gain properties in the thermodynamic limit. This approach unifies key concepts from non-Hermitian topology, including the GBZ and the winding number, and establishes a novel perspective for understanding and analyzing periodic non-Hermitian systems.

We have applied our optimization method to design novel amplifiers with a constant gain rate over a broad bandwidth, isolators with enhanced bandwidth, and frequency demultiplexers that selectively amplify signals in certain frequency ranges into different directions. Using the symbolic regression tool \textsc{AIFeynman 2} \cite{AIFeynman2,github_aifeynman}, we have automatically identified symbolic expressions for the discovered lattices, uncovering general construction rules. 
The lattices discovered with our approach are suitable for implementation in various platforms, including superconducting circuits, plasmonic waveguides, photonic crystal nanobeams, and hybrid platforms.

We envisage that in the future our method could be extended to perform tasks of increased complexity, e.g., for the discovery of disorder-robust setups, sensing, and frequency-multiplexed transport important for quantum technologies. 
Considering higher spatial dimensions, nonlinear interactions, and fermionic statistics are promising extensions that would enlarge the perspective of the field of artificial scientific discovery.\\




\begingroup
\let\MakeTextUppercase\relax
\section*{Acknowledgements}
\endgroup
\noindent We thank Christopher Eichler, Julius Gohsrich, Anton Montag, Alexander Felski, and Flore Kunst for fruitful discussions. The research is part of the Munich Quantum Valley, which is supported by the Bavarian state government with funds from the Hightech Agenda Bayern Plus. \\


\appendix

\begingroup
\let\MakeTextUppercase\relax
\section*{Methods}
\endgroup






\section{Hamiltonian and equations of motion}
\label{app:full_Hamiltonian}
We consider an open one-dimensional lattice of $N$ identical unit cells, see \cref{fig:conceptual_figure}(c). Each unit cell contains $M$ bosonic modes. When allowing for nearest-neighbor couplings between the unit cells, the most general quadratic time-independent Hamiltonian reads as:
\begin{gather}
    \hat{H} = \sum_{j=1}^N \hat{H}_j^{(0)} + \sum_{j=1}^{N-1} \hat{H}_j^{(1)} \\
    \hat{H}_j^{(0)} = \frac{1}{2} \sum_{k,l=1}^M \left( g_{kl}^{(0)} \hat{a}_{jk}^\dag \hat{a}_{jl} + \nu_{kl}^{(0)}  \hat{a}_{jk}^\dag \hat{a}_{jl}^\dag \right)
    + \mathrm{H.c.} \\
    \hat{H}_j^{(1)} \!=\! \sum_{k,l=1}^M \left( g_{kl}^{(1)} \hat{a}_{jk}^\dag \hat{a}_{j+1,l} + \nu_{kl}^{(1)}  \hat{a}_{jk}^\dag \hat{a}_{j+1,l}^\dag \right)
    + \mathrm{H.c.}
\end{gather}
Here, $\hat{a}_{jk}$ and $\hat{a}_{jk}^\dag$ denote the ladder operators of the $k$th bosonic mode in unit cell $j$. $\hat{H}_j^{(0)}$ describes all interactions within a unit cell, and $g^{(0)}$ and $\nu^{(0)}$ 
are matrices of the beamsplitter and squeezing coupling rates. Without loss generality, the coupling matrices $g^{(0)}$ can be chosen to be Hermitian and $\nu^{(0)}$ to be symmetric. The couplings between neighboring unit cells are summarized by $\hat{H}_j^{(1)}$. We note that the associated coupling matrices $g^{(1)}$ and $\nu^{(1)}$ do not possess any symmetries. As we consider identical unit cells, all coupling matrices are independent of the unit cell index.

By grouping all ladder operators of unit cell $j$ into $\hat{\xi}_j=(\hat{a}_{j1}, \hat{a}_{j2}, \ldots, \hat{a}_{j1}^\dag, \hat{a}_{j2}^\dag, ...)$, we can write $\hat{H}_j^{(0)}$ and $\hat{H}_j^{(1)}$ compactly as:
\begin{equation}
    \label{equ:def_BdG_Hamiltonians1}
    \hat{H}_j^{(0)} = \frac{1}{2} \hat{\xi}_j^\dag \hat{H}_\mathrm{BdG}^{(0)} \hat{\xi}_j, \quad 
    \hat{H}_j^{(1)} = \hat{\xi}_{j}^\dag \hat{H}_\mathrm{BdG}^{(1)} \hat{\xi}_{j+1}.
\end{equation}
Here, $\hat{H}_\mathrm{BdG}^{(0)}$ and $\hat{H}_\mathrm{BdG}^{(1)}$ express the respective Hamiltonian contributions as Bogoliubov--de-Gennes Hamiltonians and are defined as:
\begin{equation}
    \label{equ:def_BdG_Hamiltonians2}
    \hat{H}_\mathrm{BdG}^{(0)} =
    \begin{pmatrix}
        g^{(0)} & \nu^{(0)} \\
        {\nu^{(0)}}^\ast & {g^{(0)}}^\ast
    \end{pmatrix}, \text{ }
    \hat{H}_\mathrm{BdG}^{(1)} =
    \begin{pmatrix}
        g^{(1)} & \nu^{(1)} \\
        {\nu^{(1)}}^\ast & {g^{(1)}}^\ast
    \end{pmatrix}.
\end{equation}



Each mode is coupled to an external bath with $\kappa_k$ denoting the dissipation of the $k$th mode in any unit cell. 

Using input-output theory, we describe the dynamics of the open quantum system. With the considered interactions, the Langevin equations of motion equal: 
\begin{widetext}
    \begin{equation}
    \label{equ:dynamics_for_ajm}
    \dot{\hat{a}}_{jm} = - i \sum_{l=1}^M ( {g_{lm}^{(1)}}^\ast \hat{a}_{j-1,l} + \nu_{lm}^{(1)}\hat{a}_{j-1,l}^\dag + g_{ml}^{(0)}\hat{a}_{jl} + \nu_{ml}^{(0)}\hat{a}_{jl}^\dag + g_{ml}^{(1)}\hat{a}_{j+1,l} + \nu_{ml}^{(1)}\hat{a}_{j+1,l}^\dag) - \frac{\kappa_m}{2} \hat{a}_{jm} - \sqrt{\kappa_m} \hat{a}_{jm}^{(\mathrm{in})}.
    \end{equation}    
\end{widetext}
Here, $\hat{a}_{jm}^{(\mathrm{in})}$ is the input field of the external bath of the $m$th mode in the $j$th unit cell. To fulfill the OBC, we define $\hat{a}_{0,m}=\hat{a}_{N+1,m}=0$ for all $m\in[1,\ldots,M]$. 

By using \cref{equ:def_BdG_Hamiltonians1,equ:def_BdG_Hamiltonians2}, we can write \cref{equ:dynamics_for_ajm} compactly as:
\begin{align}
    \label{equ:time_evolution_xi}
    \dot{\hat{\xi}}_j =M_{-1} \hat{\xi}_{j-1} + M_0 \hat{\xi}_{j} +M_1 \hat{\xi}_{j+1} - \sqrt{K}\hat{\xi}_j^{(\mathrm{in})}.
\end{align}
Here, $M_{-1}$, $M_0$ and $M_1$ are coupling matrices of shape $2M\times 2M$ and describe the coupling to the previous, within the current and to the next unit cell. They are defined as:
\begin{align}
    &M_{-1} =-i \sigma_z {H_\mathrm{BdG}^{(1),\dag}}, \\
    &M_0 =-i \sigma_z H_\mathrm{BdG}^{(0)} - \frac{\kappa}{2}, \\
    &M_1 =-i \sigma_z H_\mathrm{BdG}^{(1)},
\end{align}
with
\begin{align}
	\sigma_z=
 \begin{pmatrix}
 \mathds{1}&0\\
 0& -\mathds{1}\\
 \end{pmatrix}.
\end{align}
Here, $\mathds{1}$ is the $M\times M$-dimensional identity.

Moreover, we introduced the diagonal matrix $K=\mathrm{diag}(\kappa_1, \ldots, \kappa_M, \kappa_1, \ldots,\kappa_M)$ and the vector $\xi_j^\mathrm{(in)}=(\hat{a}_{j1}^{(\mathrm{in})},\ldots,\hat{a}_{jM}^{(\mathrm{in})},\hat{a}_{j1}^{(\mathrm{in}),\dag},\ldots,\hat{a}_{jM}^{(\mathrm{in}),\dag})$ grouping the input signals of unit cell $j$.
Assuming that $M_1$ is invertible, we can rewrite \cref{equ:time_evolution_xi} in transfer-matrix form
\begin{equation}
    \label{equ:transfer_matrix_xi}
    \begin{pmatrix}\hat{\xi}_{j+1} \\ \hat{\xi}_j  \end{pmatrix} = T
\begin{pmatrix}\hat{\xi}_{j} \\ \hat{\xi}_{j-1}  \end{pmatrix} + \begin{pmatrix}M_1^{-1} \sqrt{K} \hat{\xi}_{j,\mathrm{in}} \\ 0\end{pmatrix},
\end{equation}
with the transfer matrix $T$ defined here as
\begin{equation}
    T = 
    \begin{pmatrix}
        - M_1^{-1} (M_0 + i \omega \mathds{1}_{2M}) & -M_1^{-1}M_{-1}\\
        \mathds{1}_{2M} & 0_{2M}
    \end{pmatrix}.
\end{equation}
Here, $\mathds{1}_{2M}$ and $0_{2M}$ denote the $2M\times 2M$ dimensional identity and zero matrix and $\omega$ the frequency in Fourier space. Note that this transfer matrix is of shape $4M \times 4M$. 

To simplify the discussion in the main text, we focus on phase-preserving devices, i.e., the system's response is independent of the quadrature of the input signal \cite{AutoScatter}. Here, the $M$ modes can be subdivided into two distinct subsets $S_1$ and $S_2$. $M_1$ modes are in subset $S_1$ and $M_2=M-M_1$ modes are in subset $S_2$. The modes within the same subset are exclusively coupled via beamsplitter couplings, while modes of different sets are coupled via two-mode squeezing. Therefore, the equations of motion for $\hat{\xi}_{1,\ldots,N}$ (see \cref{equ:time_evolution_xi,equ:transfer_matrix_xi}) break down into two decoupled sets of equations for $\hat{\vec{a}}_{1,\ldots,N}$ and $\hat{\vec{a}}_{1,\ldots,N}^\dag$. The vector $\hat{\vec{a}}_{j}$ contains the annihilation operators of all modes part of $S_1$ and the creation operators of all modes part of $S_2$. $\hat{\vec{a}}_{j}^\dag$ contains the respective conjugate operators and decouples from all $\hat{\vec{a}}_{1,\ldots,N}$. The equations for $\hat{\vec{a}}_{1,\ldots,N}$ and $\hat{\vec{a}}_{1,\ldots,N}^\dag$ are related via particle-hole symmetry \cite{AutoScatter}. 

The equations of motion for $\hat{\vec{a}}_{1,\ldots,N}$ are structurally identical to \cref{equ:time_evolution_xi}, see \cref{equ:equations_of_motion} in the main text. The only difference is that the $2M\times 2M$ matrices $M_{-1}$, $M_0$, $M_1$ and $K$ are replaced by $\mu_{-1}$, $\mu_0$, $\mu_1$ and $\kappa$, which are $M\times M$ submatrices of their more general counterparts.

We can define the transfer matrix individually for one of the decoupled sets of equations, see \cref{equ:equations_of_motion_T,equ:define_transfer_matrix}. This transfer matrix is now of size $2M \times 2 M$, simplifying our analysis. The transfer matrix of the conjugate set follows from particle-hole symmetry.

In \cref{app:derivation_gain_scaling,app:scaling_reverse_gain,app:deriviation_scaling_intermediate}, we derive the scaling of the scattering properties for the special case of phase-preserving devices. Note, however, that our derivations are more general and can easily be extended to phase-sensitive scenarios where the equations of motion do not decouple. The most notable difference is that the transfer matrix has $4M$ eigenvalues. The forward gain scales with the $(2M)$th smallest eigenvalue and the reverse gain with the inverse of the $(2M+1)$th smallest eigenvalue.





\section{Scaling of the end-to-end gain}
\label{app:derivation_gain_scaling}
In this section, we derive the scaling of the end-to-end gain in the limit $N\gg 1$. First, we diagonalize the transfer matrix $T$ (as defined in \cref{equ:define_transfer_matrix}):
\begin{equation}
    \label{equ:diagonalized_transfer_matrix}
    T = \sum_{j=1}^{2M} \lambda_j \vec{u}_j \vec{v}_j^{\,t}.
\end{equation}
Here, $\lambda_{j=1,\ldots,2M}$ are the eigenvalues of $T$ ordered according to their magnitude $\abs{\lambda_1} \leq \abs{\lambda_2} \leq \ldots \leq \abs{\lambda_{2M}}$. $\vec{u}_j$ and $\vec{v}_j$ denote the corresponding right and left eigenvectors, and $t$ the transpose.
It is important to note that $\vec{u}_j \vec{v}_j^{\,t}$ is a rank-1 matrix.


The key challenge in the derivation is the approximation of $(PT^{-N}P^t)^{-1}$ in \cref{equ:scattering_left_to_right}, as the projection operation does not commute with the inversion. To simplify the discussion, we introduce the $M\times M$-matrix $X(N)$, defined as:
\begin{align}
    \label{equ:def_X}
    X(N) &= PT^{-N}P^t = \sum_{j=1}^{2M} \lambda_j^{-N} (P\vec{u}_j) (P\vec{v}_j)^{t}.
\end{align}


Let us now focus on inverting the matrix X. 
To do so, we split $X$ into two parts:
\begin{align}
    \label{equ:def_X_0}
    X_0(N) &= \sum_{j=1}^{M^\prime} \lambda_j^{-N} (P\vec{u}_j) (P\vec{v}_j)^{t}, \\
    \label{equ:def_X_rest}
    X_\mathrm{rest}(N) &= \sum_{j=M^\prime}^{2M} \lambda_j^{-N} (P\vec{u}_j) (P\vec{v}_j)^{t}.
\end{align}
Here, $X_0$ is an invertible matrix and contains the $M^\prime$ largest summands of $X$. $M^\prime$ is chosen such that it is the smallest number required to make $X_0$ full rank. So, in general, $M^\prime\geq M$. $X_\mathrm{rest}$ contains the remaining terms. In our optimization and also the following derivations, we focus on the typical case where $M^\prime=M$. $M^\prime>M$ is only the case when special symmetries apply, e.g, when one sublattice decouples from the rest of the lattice. In such cases, we would perform our optimization anyway for the remaining lattice. 

For conciseness, we drop the dependence of $X$, $X_0$, and $X_\mathrm{rest}$ on $N$ throughout this section. Their functional dependency on $N$ will only be important in \cref{app:deriviation_scaling_intermediate}.

Using the Woodbury matrix identity, we invert $X$:
\begin{align}
    \label{equ:Woodbury_identity}
    X^{-1} = \left[\sum_{n=0}^\infty (- X_0^{-1} X_\mathrm{rest})^n \right]X_0^{-1}.
\end{align}
The largest term in $X_0^{-1}$ scales with $\lambda_{M}^N$, in $X_\mathrm{rest}$ the largest term scales with $\lambda_{M+1}^{-N}$. Therefore, the dominant contribution in the sum comes from $n=0$; all other terms with $n>0$ will decay exponentially in the thermodynamic limit, so we can approximate $X^{-1}$ as:
\begin{equation}
    \label{equ:scaling_X_inv}
    X^{-1} = \left[\mathds{1}+\mathcal{O}\left(\frac{\lambda_{M}^N}{\lambda_{M+1}^N}\right)\right] X_0^{-1}.
\end{equation}
So, in leading order, $X^{-1}$ and also the end-to-end scattering matrix $S_\mathrm{RL}(\omega)$, describing the amplification from left to right, scale with the $M$th smallest eigenvalue $\lambda_{M}^N$.
With the symbol $\mathcal{O}$, we indicate the next higher-order corrections to the respective terms. When multiple arguments appear, each represents a possible higher-order contribution, any of which may dominate depending on the specific circumstances.


To derive the prefactor of this scaling, we write $X_0$ as:
\begin{equation}
    \label{equ:X_0_decomposition}
    X_0(N) = U \Lambda(N) V^t.
\end{equation}
$U$ and $V$ are $M\times M$ matrices and are defined as $U=(P\vec{u}_1, \ldots, P\vec{u}_M)$ and $V=(P\vec{v}_1, \ldots, P\vec{v}_M)$. $\Lambda$ is a diagonal matrix and equals $\mathrm{diag}(\lambda_1^{-N}, \ldots, \lambda_M^{-N})$. Note that $U^{-1}\neq V$, so the vectors $P\vec{u}_j$ and $P\vec{v}_j$ are not eigenvectors of $X_0$.

In leading order, we can approximate $X_0^{-1}$ as:
\begin{equation}
    X_0^{-1} = \lambda_M^N (V^t)^{-1} \mathrm{diag}(0,\ldots,0,1) U^{-1}  + \mathcal{O} (\lambda_{M-1}^N).
\end{equation}


Substituting everything back into \cref{equ:scattering_left_to_right}, we find the ideal gain properties in the limit of large $N$:
\begin{equation}
    S_\mathrm{RL}(\omega) = A \lambda_M^N \left[ \mathds{1}  + \mathcal{O}\left(\frac{\lambda_{M-1}^N}{\lambda_{M}^N}, \frac{\lambda_{M}^N}{\lambda_{M+1}^N}\right) \right].
\end{equation}
Here, $A$ is the prefactor of this scaling and equals:
\begin{equation}
    \label{equ:scaling_prefactor}
    A = -\sqrt{\kappa} (V^t)^{-1} \mathrm{diag}(0,\ldots,0,1) U^{-1} \mu_{-1}^{-1}\sqrt{\kappa}.
\end{equation}

The corrections to this scaling law decay exponentially with $\lambda_{M-1}/\lambda_M$ and $\lambda_{M}/\lambda_{M+1}$. Therefore, when other eigenvalues have an absolute value close to $\abs{\lambda_M}$, they interfere with the leading order term. The eigenvalues, and importantly their phases, depend on the frequency $\omega$. So, whether the interference is constructive or destructive is frequency dependent. This gives rise to fluctuations as a function of the frequency, which are typically labeled as ripples, see \cref{fig:gain_profile}(a). The strength of the ripples compared to the gain decays exponentially with the lattice size. However, their speed of decay is determined by $\abs{\lambda_{M-1}/\lambda_M}$, or $\abs{\lambda_{M}/\lambda_{M+1}}$, depending on which term is larger. Therefore, we characterize the strength of the ripples over those ratios. \\

\section{Scaling of the reverse gain}
\label{app:scaling_reverse_gain}
In this section, we derive the scattering matrix elements describing the reverse gain, so the scattering from the last to the first unit cell. To do so, we repeat the derivation of \cref{equ:scattering_left_to_right}, just in the reverse direction.

Starting from $j=N$, we write \cref{equ:equations_of_motion_T} as:
\begin{equation}
    \begin{pmatrix}0 \\ \vec{a}_N  \end{pmatrix} = T
\begin{pmatrix}\vec{a}_{N} \\ \vec{a}_{N-1}  \end{pmatrix} + \begin{pmatrix}\mu_1^{-1} \sqrt{\kappa} \vec{a}_{N,\mathrm{in}} \\ 0\end{pmatrix}.
\end{equation}
We now iteratively insert \cref{equ:equations_of_motion_T} for decreasing $j$. For simplicity, we assume that all input signals for $j\neq N$ are zero. At the leftmost unit cell $j=1$, we get:
\begin{equation}
    \begin{pmatrix}0 \\ \vec{a}_N  \end{pmatrix} = T^N
\begin{pmatrix}\vec{a}_{1} \\ 0  \end{pmatrix} + \begin{pmatrix}\mu_1^{-1} \sqrt{\kappa} \vec{a}_{N,\mathrm{in}} \\ 0\end{pmatrix}.
\end{equation}

Solving for $\vec{a}_1$ and using the boundary condition, we calculate the output field $\vec{a}_{1,\mathrm{out}}$:
\begin{equation}
    \vec{a}_{1,\mathrm{out}} = S_\mathrm{LR} \vec{a}_{N,\mathrm{in}} + \vec{a}_{1,\mathrm{in}}.
\end{equation}
Here, we defined the scattering matrix for the reverse direction $S_\mathrm{LR}$ as:
\begin{equation}
    \label{equ:scattering_matrix_reverse}
    S_\mathrm{LR}(\omega) = -\sqrt{\kappa}(P_\mathrm{rev}T^{N}P_\mathrm{rev}^t)^{-1}\mu_{1}^{-1}\sqrt{\kappa},
\end{equation}
with the $M\times 2M$ matrix $P_\mathrm{rev}=(\mathds{1}, 0)$, where $P_\mathrm{rev} \bullet P_\mathrm{rev}^t$ selects the upper left block of a matrix $\bullet$. 

The reverse scattering matrix is directly connected to the forward scattering in \cref{equ:scattering_left_to_right} by symmetry. The transfer matrix $T$ describes the forward propagation of excitations along the lattice, while the inverse $T^{-1}$ describes the propagation in the reverse direction; therefore, replacing $T$ when comparing \cref{equ:scattering_left_to_right} to \cref{equ:scattering_matrix_reverse}. This has important implications for the idealized scaling of the reverse gain: While the eigenvalues of the transfer matrix determine the forward gain with $\lambda_M$ providing the dominant contribution, the reverse gain is determined by the inverse of the eigenvalues with $1/\lambda_{M+1}$ as the dominant contribution.

It is straightforward to repeat the approximation in \cref{app:derivation_gain_scaling}. The most important difference is that we sort all terms of $X_\mathrm{rev}=P_\mathrm{rev}T^{N}P_\mathrm{rev}^t$ from the largest contribution $\lambda_{2M}^N$ to the smallest one $\lambda_{1}^N$:
\begin{equation}
    X_\mathrm{rev}=\sum_{j=1}^{2M} \lambda_j^N (P_\mathrm{rev} \vec{u}_j)(P_\mathrm{rev} \vec{v}_j)^t.
\end{equation}
Grouping the largest $M$ terms into the matrix $X_0$, and using the Woodbury matrix identity, we obtain:
\begin{equation}
    S_\mathrm{LR}(\omega) = A_\mathrm{rev} \lambda_{M+1}^{-N}\left[ \mathds{1}  + \mathcal{O}\left(\frac{\lambda_{M-1}^N}{\lambda_{M}^N}, \frac{\lambda_{M}^N}{\lambda_{M+1}^N}\right) \right].
\end{equation}
Here, $A_\mathrm{rev}$ is the prefactor of the scaling and equals:
\begin{equation}
    A_\mathrm{rev} = -\sqrt{\kappa} (V_\mathrm{rev}^t)^{-1} \mathrm{diag}(0,\ldots,0,1) U_\mathrm{rev}^{-1} \mu_{1}^{-1}\sqrt{\kappa}.
\end{equation}
with $U_\mathrm{rev}=(P_\mathrm{rev}\vec{u}_{2M}, \ldots, P_\mathrm{rev}\vec{u}_{M+1})$ and $V_\mathrm{rev}=(P_\mathrm{rev}\vec{v}_{2M}, \ldots, P_\mathrm{rev}\vec{v}_{M+1})$ being defined in analogy to the scaling for the forward gain, see \cref{app:derivation_gain_scaling}. \\

\section{Scattering from intermediate unit cells}
\label{app:deriviation_scaling_intermediate}
In the main text and \cref{app:derivation_gain_scaling,app:scaling_reverse_gain}, we derived the exact and approximate scattering from one end of the lattice to the other. However, for many applications, it is essential to optimize the scattering from intermediate unit cells to the ends. For example, consider a lattice acting as an amplifier. Typically, noise leaks in at intermediate unit cells and is amplified together with the signal of interest. Furthermore, we consider in the main text the example of a frequency demultiplexer. Here, the signal of interest is injected in the middle of the lattice and is scattered to the respective ends. Therefore, in this section, we repeat the analysis of \cref{app:derivation_gain_scaling,app:scaling_reverse_gain} for those scattering paths.

To anticipate the main result of this section: We show below that the gain for any scattering path from an intermediate unit cell to the right or left end still scales with $\lambda_M$ or $1/\lambda_{M+1}$ to the power of the number of passed unit cells. However, we note that the scaling prefactor changes compared to the scattering from one end to the other.

First, we analyse the forward direction and start from \cref{equ:derive_forward_scattering_j_1}. We iteratively insert \cref{equ:equations_of_motion_T}. However, this time we explicitly consider input signals at intermediate unit cells. In this case, \cref{equ:to_solve_for_gain} generalizes to:
\begin{equation}
    \label{equ:to_solve_for_gain_intermediate}
    \begin{pmatrix} \vec{a}_1 \\ 0  \end{pmatrix} = 
T^{-N}
\begin{pmatrix}0 \\ \vec{a}_{N} \end{pmatrix} +\sum_{j=1}^N T^{-(j-1)} \begin{pmatrix} 0\\\mu_{-1}^{-1} \sqrt{\kappa} \vec{a}_{j,\mathrm{in}} \end{pmatrix}.
\end{equation}
By solving \cref{equ:to_solve_for_gain_intermediate} and using the boundary conditions, we get:
\begin{equation}
    \vec{a}_{N,\mathrm{out}} =\vec{a}_{1,\mathrm{in}} + \sum_{j=1}^N S_{\mathrm{R}j}(\omega) \vec{a}_{j,\mathrm{in}},
\end{equation}
with $S_{\mathrm{R}j}(\omega)$ being defined as:
\begin{equation}
    \label{equ:S_j_forward}
    S_{\mathrm{R}j}(\omega) = -\sqrt{\kappa}(PT^{-N}P^t)^{-1} (PT^{-(j-1)}P^t)  \mu_{-1}^{-1}\sqrt{\kappa}.
\end{equation}
$S_{\mathrm{R}j}(\omega)$ is a scattering matrix describing the scattering from unit cell $j$ to the last unit cell (the right end of the lattice).

With the definitions in \cref{equ:def_X,equ:def_X_0,equ:def_X_rest} and for $1<j<N$, we rewrite \cref{equ:S_j_forward} as:
\begin{widetext}
    \begin{subequations}
        \begin{align}
            S_{\mathrm{R}j}(\omega) =& -\sqrt{\kappa}\left[X_0(N) + X_\mathrm{rest}(N)\right]^{-1} \left[ X_0(j-1) + X_\mathrm{rest}(j-1) \right]  \mu_{-1}^{-1}\sqrt{\kappa} =\\
            \label{equ:derivation_S_j_b}
                =& -\sqrt{\kappa} \left[\mathds{1}+\mathcal{O}\left(\frac{\lambda^N_{M}}{\lambda^N_{M+1}}\right)\right] X_0(N)^{-1} \left[ X_0(j-1) + X_\mathrm{rest}(j-1) \right]  \mu_{-1}^{-1}\sqrt{\kappa} =\\
                \label{equ:derivation_S_j_c}
            = & -\sqrt{\kappa} \left[\mathds{1}+\mathcal{O}\left(\frac{\lambda^N_{M}}{\lambda^N_{M+1}}\right)\right]  \left[ (V^t)^{-1}\Lambda(-N) U^{-1} U \Lambda(j-1) V^t + X_0(N)^{-1} X_\mathrm{rest}(j-1) \right]  \mu_{-1}^{-1}\sqrt{\kappa} =\\
            \label{equ:derivation_S_j_d}
            = & -\sqrt{\kappa} \left[\mathds{1}+\mathcal{O}\left(\frac{\lambda^N_{M}}{\lambda^N_{M+1}}\right)\right]  \left[ (V^t)^{-1}\Lambda(j-1-N) V^t + X_0(N)^{-1} X_\mathrm{rest}(j-1) \right]  \mu_{-1}^{-1}\sqrt{\kappa} =\\
            =& -\sqrt{\kappa} \left[\mathds{1}+\mathcal{O}\left(\frac{\lambda^N_{M}}{\lambda^N_{M+1}}\right)\right]  \left[ (V^t)^{-1}\Lambda(j-1-N) V^t + \mathcal{O}\left(\frac{\lambda^N_{M}}{\lambda^{j-1}_{M+1}}\right) \right]  \mu_{-1}^{-1}\sqrt{\kappa} =\\
            =& -\sqrt{\kappa} \left[\mathds{1}+\mathcal{O}\left(\frac{\lambda^N_{M}}{\lambda^N_{M+1}}\right)\right]  \left[ (V^t)^{-1}\mathrm{diag}(0,\ldots,0,1) \lambda_M^{N-j+1} V^t + \mathcal{O}\left( \lambda_{M-1}^{N-j+1},\frac{\lambda^N_{M}}{\lambda^{j-1}_{M+1}} \right) \right]  \mu_{-1}^{-1}\sqrt{\kappa} =\\
            =& \underbrace{-\sqrt{\kappa} (V^t)^{-1}\mathrm{diag}(0,\ldots,0,1) V^t \mu_{-1}^{-1}\sqrt{\kappa}}_{=A_{\mathrm{RC}}} \lambda_M^{N-j+1} \left[\mathds{1} + \mathcal{O}\left(\frac{\lambda^N_{M}}{\lambda^N_{M+1}},\frac{\lambda_{M-1}^{N-j+1}}{\lambda_M^{N-j+1}},\frac{\lambda^{j-1}_{M}}{\lambda^{j-1}_{M+1}}\right) \right].
        \end{align}
    \end{subequations}    
\end{widetext}

In \cref{equ:derivation_S_j_b}, we use the Woodbury-matrix identity to approximate the inverse of $X(N)$, see also \cref{app:derivation_gain_scaling}. In \cref{equ:derivation_S_j_c}, we drag $X_0(N)^{-1}$ into the right bracket, and expand $X_0$ in the first summand using \cref{equ:X_0_decomposition}. In \cref{equ:derivation_S_j_d} we use $\Lambda(x)+\Lambda(y)=\Lambda(x+y)$, and determine the leading order term for the second summand. Here, $X_0(N)^{-1}$ is dominated by $\lambda_M^N$, and $X_\mathrm{rest}(j-1)$ by $\lambda_{M+1}^{-(j-1)}$. In the next line, we approximate the first summand accordingly. After sorting all terms, we find that $S_{\mathrm{R}j}$ scales with $\lambda_M$ to the power of the number of passed unit cells, as one would have expected intuitively. However, the prefactor of the scaling changed, see \cref{equ:scaling_prefactor} for comparison.

Higher order corrections to this scaling vanish for $N\to \infty$ and $1 \ll j \ll N$. So, this scaling is well fulfilled for unit cells that are sufficiently distanced from the boundaries.

It is straightforward to repeat these steps for the scattering from intermediate unit cells to the left.
Therefore, we will only quickly summarize the results here. Considering input signals from all unit cells, the output field at the first unit cell turns out to equal:
\begin{equation}
    \vec{a}_{1,\mathrm{out}} =\vec{a}_{1,\mathrm{in}} + \sum_{j=1}^N S_{\mathrm{L}j}(\omega) \vec{a}_{j,\mathrm{in}},
\end{equation}
with
\begin{equation}
    S_{\mathrm{L}j} = - \sqrt{\kappa}(P_\mathrm{rev}T^{N}P_\mathrm{rev}^t)^{-1} (P_\mathrm{rev}T^{N-j}P_\mathrm{rev}^t)  \mu_{1}^{-1}\sqrt{\kappa}.
\end{equation}
Here, $S_{\mathrm{L}j}(\omega)$ describes the scattering from unit cell $j$ to the left end of the lattice. In the thermodynamic limit and for $1\ll j \ll N$, $S_{\mathrm{L}j}(\omega)$ scales according to:
\begin{equation}
    S_{\mathrm{L}j}(\omega) \approx A_\mathrm{LC} \lambda_{M+1}^{-j}.
\end{equation}
Here, $A_{\mathrm{LC}}$ is the prefactor of this scaling and equals:
\begin{equation}
    A_\mathrm{LC} = -\sqrt{\kappa} (V_\mathrm{rev}^t)^{-1}\mathrm{diag}(0,\ldots,0,1) V_\mathrm{rev}^t \mu_{1}^{-1}\sqrt{\kappa}.
\end{equation}

\section{Dynamical matrix and stability under OBC}
\label{app:stability}
The equations of motion in \cref{equ:equations_of_motion} can describe unstable motions drifting away from the saddle points. To determine whether a system is stable, we have to calculate the OBC spectrum. To do so, we combine the equations~\eqref{equ:equations_of_motion} for all unit cells to a single equation describing the time evolution of the full system:
\begin{equation}
    \dot{\vec{A}} = -i H \vec{A} - \sqrt{\mathcal{K}} \vec{A}_\mathrm{in}.
\end{equation}
Here, the vector $\vec{A}=(\vec{a}_1,\ldots,\vec{a}_N)$ contains all degrees of freedom of the lattice. $\mathcal{K}$ is a block-diagional matrix with $N\times N$ blocks. All of its diagonal blocks equal the matrix $\kappa$. $H$ is the non-Hermitian dynamical matrix of the full system and contains $N\times N$ blocks, each of the shape $M \times M$. As we consider an open lattice with nearest-neighbor couplings, $H$ is a block-tridiagonal Toeplitz matrix of the form: 
\begin{equation}
    \label{equ:full_dynamical_matrix}
    H = \begin{pmatrix}
        i\mu_0 & i\mu_1 & 0 & &\ldots & 0 \\
        i\mu_{-1} & i\mu_0 & i\mu_1 & 0 &  & \\
        0 & i\mu_{-1} & i\mu_0 & i\mu_1 &  & \vdots \\
        \vdots & & \ddots & \ddots& \ddots & 0 \\
         &  & &   i\mu_{-1} & i\mu_0 & i\mu_1 \\
        0 & \ldots & & 0  & i\mu_{-1} & i\mu_0
    \end{pmatrix}.
\end{equation}
The system's motion is stable and converges to a stable steady state as long as the dynamical matrix $H$ has only eigenvalues with a negative imaginary part. 



\section{Invertibility of the coupling matrices}
\label{app:invertiblity_problem}

For the construction of the transfer matrix, we assumed that the coupling matrices $\mu_{\pm 1}$ are invertible, see \cref{equ:define_transfer_matrix}. However, this is not always the case. Especially, when the number of links between adjacent unit cells is smaller than $M$, $\mu_{\pm 1}$ is never invertible. To overcome this problem when computing the transfer matrix spectrum numerically in our optimization, we add a small $\epsilon=\num{1e-8}$ on the diagonal of $\mu_{\pm 1}$. 
This allows us to compute the true scaling with reasonable precision and to optimize lattices where $\mu_{\pm 1}$ is not invertible. E.g., lattice (I) in \cref{fig:isolator}(c) has non-invertible $\mu_{\pm 1}$, and is nevertheless discovered by our optimization scheme.

However, this approach has its limits. The determinant of $(\mu_{\pm 1} + \epsilon \mathds{1})$ scales with $\epsilon^{\mathrm{M - rank(\mu_{\pm 1})}}$. In our work, we focused mostly on $M=2$, so the difference between $M$ and the rank is at most \num{1} before the unit cells get disconnected. 
The inverse matrix of $(\mu_{\pm 1} + \epsilon \mathds{1})$ cannot be computed anymore with reasonable numerical precision, when the deviation between $M$ and the rank becomes too large when considering larger $M$. 

To overcome this challenge in future extensions, it might be helpful to use the definition of the transfer matrix introduced in \cite{dwivedi2016bulk}. Here, the rank of the coupling matrices 
is interpreted as the number of links between two adjacent unit cells when choosing a mode basis diagonalizing the coupling matrices. If the rank of the coupling matrices is smaller than $M$, some modes (within this basis) are disconnected. The central idea presented in \cite{dwivedi2016bulk} is to map out those disconnected modes and to define the transfer matrix for the reduced subspace, thereby avoiding the inversion problem.



\section{Connection between the transfer matrix and topology and other concepts}
\label{app:connection_to_topology}


The transfer matrix is closely connected to concepts from non-Hermitian topology and unifies existing approaches for analyzing non-Hermitian systems. In particular, the eigenvalues of the transfer matrix are directly related to the GBZ and winding number. In the following, we explain those concepts and show the relation between them and the transfer-matrix approach.



Following \cite{yokomizo2019non}, we apply the generalized Bloch ansatz $\vec{a}_j (t) = \vec{v} e^{i (kj - Et) }$. Here, $E$ is the energy (an eigenvalue of the dynamic matrix $H$), which is in general complex, as the considered systems are non-Hermitian. Its real part is the oscillation frequency, and its imaginary part describes whether an excitation is decaying or increasing over time. $k$ is the wave vector. For PBC, $k$ is real and describes solutions that are periodic in space. For OBC, $k$ is complex, and its imaginary part describes solutions that decay or rise along the lattice.

Inserting this ansatz into \cref{equ:equations_of_motion} and setting any input field $\vec{a}_{j,\mathrm{in}}$ to zero, gives the eigenvalue problem:
\begin{equation}
    \label{equ:EV_problem_PBC}
    (H(k) - E\mathds{1}) \vec{v} = 0.
\end{equation}
This eigenvalue problem has the characteristic polynomial:
\begin{equation}
    \label{equ:characteristic_polynomial}
    \det (H(k) - E\mathds{1}) = 0.
\end{equation}
Here, $H(k)$ is the Bloch Hamiltonian and equals
\begin{equation}
    H(k) = i\mu_{-1} e^{-\mathrm{i}k} + i\mu_0 + i\mu_1 e^{\mathrm{i}k}
\end{equation}
for the kind of lattices defined in the main text. The corresponding characteristic polynomial equals:
\begin{equation}
    \label{equ:our_characteristic_polynomial}
    \det (\mu_{-1}\beta^{-1} + (\mu_0 + iE\mathds{1}) + \mu_1 \beta) = 0.
\end{equation}
Here, we introduced the short-hand notation $\beta=\exp(ik)$, simplifying the following discussions.

\cref{equ:characteristic_polynomial} defines the spectrum in the thermodynamic limit. Any $E$ fulfilling \cref{equ:characteristic_polynomial} for a real $k\in [-\pi,\pi]$ is part of the PBC spectrum. The corresponding $\beta$ form a unit circle in the complex plane. This circle is called the Brillouin Zone (BZ) \cite{yokomizo2019non}.

The GBZ extends this concept to OBC. As shown in~\cite{Yao2018edge,yokomizo2019non} and already discussed in the main text, the construction of the GBZ involves solving \cref{equ:characteristic_polynomial} for $\beta(E)$ and sorting the roots according to their magnitude, so $\abs{\beta_1}\leq \ldots \leq \abs{\beta_{2M}}$. Any $E$ fulfilling $\abs{\beta_M}=\abs{\beta_{M+1}}$ is part of the OBC spectrum. This condition arises from the requirement that the eigenvectors $\vec{v}$ have to satisfy the OBC in the thermodynamic limit.

The transfer matrix is closely related to those concepts. To show this, let us write down the eigenvalue problem of the transfer matrix defined in \cref{equ:define_transfer_matrix}:
\begin{subequations}
\begin{align}
        &\det (T - \mathds{1}_{2M}\lambda) = \\
        &=\det (\mu_1^{-1}\mu_{-1} + \mu_1^{-1}(\mu_0 + i\omega \mathds{1}) \lambda + \mathds{1} \lambda^2) =\\
        &=\det (\mu_1^{-1})\lambda \det (\mu_{-1} \lambda^{-1} + (\mu_0 + i\omega \mathds{1}) + \mu_1 \lambda).
\end{align}
\end{subequations}
As shown, the eigenvalues of the transfer matrix are roots of the characteristic polynomial in \cref{equ:our_characteristic_polynomial} after making the substitutions $\beta=\lambda$ and $E=\omega$.
This link is well known for Hermitian systems, where the transfer matrix is used to calculate the system's spectral properties \cite{chang1982complex,lee1981simple}.

Due to the equivalence between the eigenvalues $\lambda$ of the transfer matrix and the roots $\beta$ of the characteristic polynomial, we can draw multiple connections to other concepts of non-Hermitian topology. In the context of the GBZ, it was shown that the forward end-to-end gain scales exponentially with $\abs{\beta_M(\omega)}$, while the reverse gain scales with $1/\abs{\beta_{M+1}(\omega)}$ \cite{xue2021simple}. This agrees directly with our derivations based on the transfer matrix, Eq.~\eqref{equ:scattering_approximate}. 

Furthermore, the equivalence of $\lambda$ and $\beta$ establishes a direct connection between the transfer matrix and the winding number defined in \cref{equ:winding_number}. Due to the argument principle, the winding number equals the number of zeros minus the number of poles of $\det (H(\beta) - \omega \mathds{1})$ enclosed in a circle with $\abs{\beta}=1$, i.e., the BZ \cite{zhang2020correspondence}. The number of poles is always fixed to $M$, see \cref{equ:our_characteristic_polynomial}. The number of zeros enclosed within $\abs{\beta}=1$ is the number of transfer-matrix eigenvalues with $\abs{\lambda_j}<1$, leading to \cref{eq:winding}.

\section{Details on the discrete optimization}

We explore the discrete search space of all possible lattices using a brute-force search, i.e., every lattice layout is tested individually by the continuous optimization. As a result, our approach identifies all lattice layouts within the search space that yield the target behavior.

In our current work, we restricted our automated discovery to up to two modes per unit cell ($M=2$), resulting in a discrete search space containing only up to 500 different lattice layouts. 
Due to this small number, the brute-force search can be effectively parallelized, yielding faster results than any other algorithm aiming to characterize all lattices in the search space. 

As the search space grows super-exponentially with $M$, a brute-force search is not viable for more than two modes per unit cell. For future extensions with larger $M$, we recommend using the discrete optimization of {\sc AutoScatter}, which employs an intelligent breadth-first search combined with graph theory to investigate the search space, see \cite{AutoScatter,github_autoscattering} for more details.



\section{Details on the continuous optimization}
\label{app:details_continuous_optimization}

We use particle swarm optimization (PSO) \cite{engelbrecht2007computational} to find a suitable parameter set $\vec{x}^{\,\ast}$ that minimizes the loss function defined in \cref{equ:loss_function}. 
PSO is a gradient-free method and uses a swarm of particles, where each particle $i$ sits at a position $\vec{x}_i$ in parameter space and has a velocity $\vec{v}_i$. The mass of each particle equals $w_P$. The particles follow simple force fields, where each particle is attracted by the position with the smallest loss value it has seen during its movement, and by the best position any particle has seen so far. The strength of these force fields is controlled by the hyperparameters $c_1$ and $c_2$, respectively, see \cite{engelbrecht2007computational} for more details. In our work, we use the global-best PSO algorithm of the Python library {\sc PySwarms} \cite{pyswarmsJOSS2018}. For the optimization, we set the hyperparameters to $c_1=0.2$, $c_2=0.9$, and $w_P=0.9$, which we found to yield the best performance. We set the particle number to $50$, providing a good trade-off between performance and computational cost.

In our previous work {\sc AutoScatter} \cite{AutoScatter,github_autoscattering} we employed a gradient-based approach for the continuous optimization. For our current work, we decided against this because the loss function is not steady or even undefined for some points in parameter space. In all examples discussed in the main text, we enforce that the first or higher order derivatives of $\abs{\lambda_M}$ with respect to $\omega$ equal zero. However, the order of the eigenvalues can change. When such a transition occurs, the derivatives of $\lambda_M$ with respect to $\omega$ make a sudden jump, and at the transition point, they are not even defined. We found that our previously employed gradient-based approaches easily get stuck in this highly non-convex and unsteady loss landscape. On the other hand, we found that PSO still performs decently. Even if a handful of particles get stuck, the other particles continue their dynamic, and the attraction to the globally best seen position can dislodge the stuck particles.

Our approach allows for combining multiple optimization targets, see \cref{app:optimization_constraints} for a complete list and more information on their implementation. Each of those optimization targets is expressed by a function $f(\vec{x})$, where $\vec{x}$ denotes all free continuous parameters of the considered lattice layout. We only consider optimization targets that can be fulfilled exactly, so there exists at least one parameter set $\vec{x}^\ast$, such that $\abs{f(\vec{x}^\ast)}^2=0$. 

In a multiobjective optimization problem, we only label candidate lattices as valid when they can fulfill all targets perfectly. This means that we can define the loss function $\mathcal{L}$ simply as the quadratic sum of all targets, see \cref{equ:loss_function}. More complicated definitions, such as a weighted quadratic sum, might improve convergence speed or reduce the number of local minima, but will not alter the global minima, where $\mathcal{L}$ equals zero.

The calculation of all optimization targets and the loss function is implemented using the Python library {\sc Jax}. 





\section{Optimization targets}
\label{app:optimization_constraints}

In summary, our approach covers the following optimization targets:
\begin{enumerate}
    \item enforce a target value for the gain rate per unit cell at some target frequency
    \item enforce a target value for the scaling prefactor at some target frequency 
    \item enforce a target value (typically zero) for any of their derivatives at some target frequency
    \item enforce a minimum distance between two eigenvalues for a range of frequencies
    \item enforce dynamic stability for an open lattice with a defined length
\end{enumerate}

The first four optimization targets are defined over the eigenspectrum of the transfer matrix, which is generally a non-Hermitian matrix. In the latest version of {\sc Jax} (v0.6.2), the computation of non-Hermitian eigenspectra is not supported on GPU, thereby restricting us to CPU usage. Furthermore, {\sc Jax} does currently not support the calculation of any derivatives of eigenvectors or higher-order derivatives of the eigenvalues using autodifferentiation. To implement constraints of type 3, we approximate any of the required derivatives using finite differences. 
We note that constraints 1 to 3 are also available for the reverse gain and for scattering from one of the central unit cells to any of the ends.

Excluding the stability constraint, all optimization targets are defined over the transfer matrix and, therefore, describe the idealized gain properties in the thermodynamic limit. 
Constraint 5 only enforces that an open lattice of fixed length (typically set to \num{10}) is dynamically stable as defined in \cref{app:stability}. This does not guarantee that the lattice is stable in the thermodynamic limit. However, in combination with constraint 4, this can be guaranteed. 

As discussed in the main text, if the OBC spectrum in the thermodynamic limit has a band that crosses or touches the imaginary axis, the lattice is either unstable or on the verge of instability. 
This is prevented by constraint 4, whose original purpose was to establish a minimum distance between the eigenvalues to reduce the ripple strength.

With constraint 4 in place, there is still the possibility that a separate band of the OBC spectrum in the thermodynamic limit is entirely above the imaginary axis. 
Assuming that the OBC spectrum for the finite lattice tested with constraint 5 has already loosely converged towards its thermodynamic limit, this situation can also be excluded, and the lattice is guaranteed to be stable in the thermodynamic limit.

\bibliography{bibi}
\end{document}


\title{Supplemental Material for ``Artificial discovery of lattice models for wave transport''}

\author{Jonas~Landgraf}\email{Jonas.Landgraf@mpl.mpg.de}
\affiliation{Max Planck Institute for the Science of Light, Staudtstr.~2, 91058 Erlangen, Germany}
\affiliation{Physics Department, University of Erlangen-Nuremberg, Staudstr.~5, 91058 Erlangen, Germany}
\author{Clara~Wanjura}
\affiliation{Max Planck Institute for the Science of Light, Staudtstr.~2, 91058 Erlangen, Germany}
\author{Vittorio~Peano}
\affiliation{Max Planck Institute for the Science of Light, Staudtstr.~2, 91058 Erlangen, Germany}
\author{Florian~Marquardt}
\affiliation{Max Planck Institute for the Science of Light, Staudtstr.~2, 91058 Erlangen, Germany}
\affiliation{Physics Department, University of Erlangen-Nuremberg, Staudstr.~5, 91058 Erlangen, Germany}

\date{\today}

\maketitle

\begin{appendix}
\onecolumngrid
\renewcommand\thefigure{S\arabic{figure}}
\renewcommand\thetable{S\arabic{table}}

\begin{figure*}
    \centering
    \includegraphics[width=.9\linewidth]{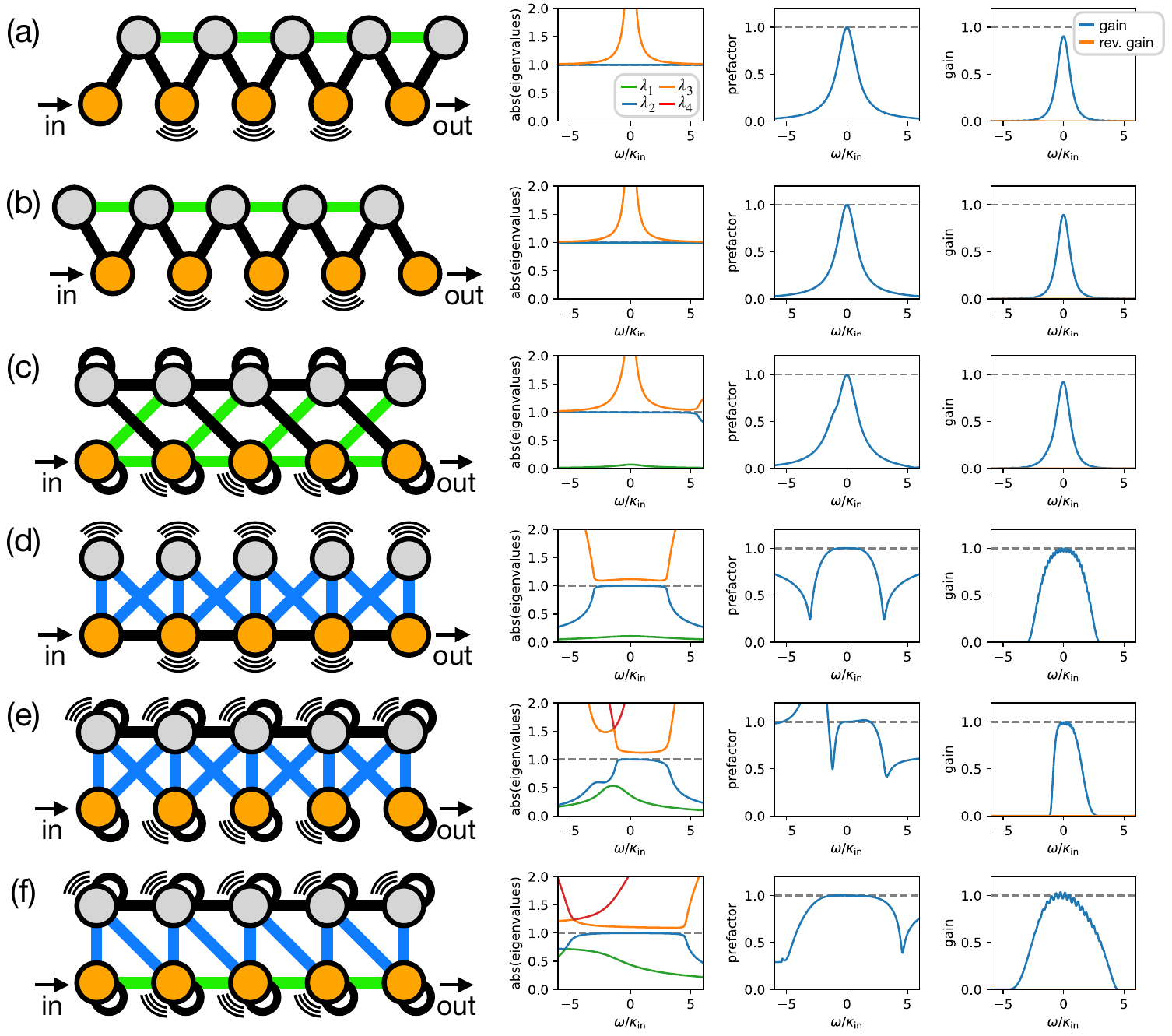}
    \caption{Summary of all irreducible isolator chains discovered by our method with the target properties described in the main text. For each chain, we show the transfer-matrix spectrum, the scaling prefactor, and the gain and reverse gain for a chain length of $N=40$ as a function of the frequency $\omega$.}
    \label{fig:all_isolators}
\end{figure*}

\begin{figure*}
    \centering
    \includegraphics[width=.9\linewidth]{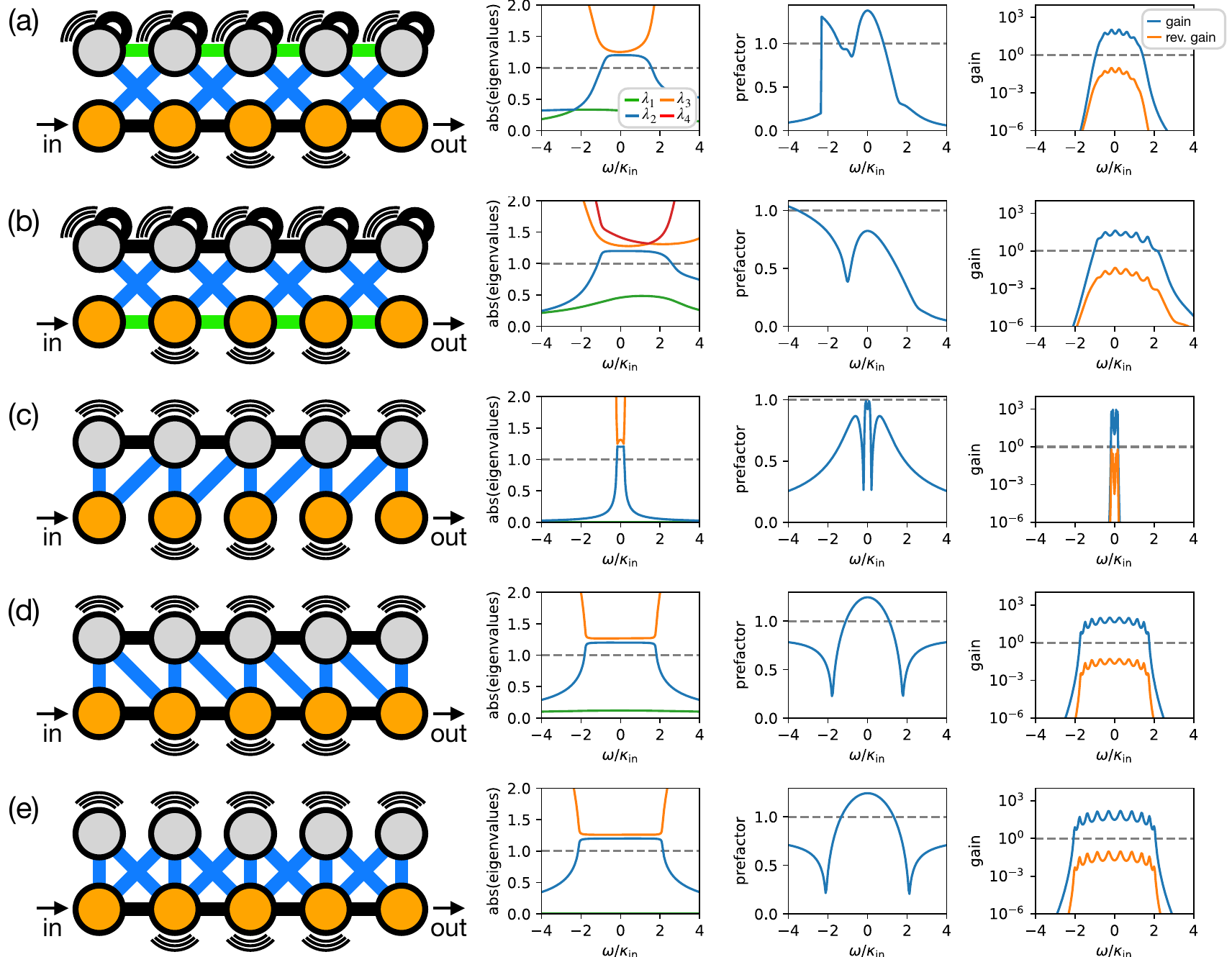}
    \caption{Summary of all irreducible amplifier chains discovered by our method with the target properties described in the main text. For each chain, we show the transfer-matrix spectrum, the scaling prefactor, and the gain and reverse gain for a chain length of $N=10$ as a function of the frequency $\omega$.}
    \label{fig:all_amplifiers}
\end{figure*}

\section{Summary of all discovered isolator and amplifier chains}

In \cref{fig:all_isolators}, we give a quick summary of all isolators discovered by our method with the target properties described in the main text. All shown isolators are irreducible, meaning that none of their coupling rates or phases can be set to zero without violating the target properties. For each chain, we show its idealized gain properties exemplarily for a single parameter set discovered by our method. 

Of the discovered lattices, (a--c) only use beamsplitter couplings, while (d--f) also make use of squeezing. 
For the lattices (a--c), the gain rate $\abs{\lambda_M}$ is perfectly constant over the whole observed frequency window. However, the bandwidth is limited by the shape of the gain prefactor, which has a non-vanishing second derivative at $\omega=0$. In contrast, for the lattices (d--f), we were able to enforce that the second derivative of the prefactor vanishes. For lattice (f), even the third derivative vanishes, giving it the best bandwidth of the discovered lattices. The lattices (b) and (f) are also discussed in the main text.

We note that the transmission of the isolators (a--c) in \cref{fig:all_isolators} is slightly below \num{1}. This discrepancy arises from a small deviation of $\abs{\lambda_M(\omega=0)}$ from its target value $\num{1}$, specifically by around of \num{0.002}. This difference is exponentially amplified with increasing chain length, leading to a noticeable deviation in the gain profile for a chain length of \num{40}.

In \cref{fig:all_amplifiers}, we summarize all irreducible amplifiers that fulfill the target properties described in the main text. The amplifiers (d) and (e) have a constant gain rate profile within their bandwidth. Furthermore, we discovered the amplifiers (a) and (b), where the gain rate is flat, but not to the same degree as for (d) and (e). 

Lastly, our algorithm also discovered the lattice (c), which illustrates a bottleneck of our approach. For our algorithm, we demand that the gain rate, the gain prefactor, and their derivatives fulfill certain constraints, but typically only at one specific target frequency. Lattice (c) fulfills those characteristics, but not for points close to this target frequency. Therefore, its bandwidth is small, the gain varies abruptly, and it is generally not a well-working amplifier. Luckily, we discovered only a small number of lattices where the frequency dependence close to the target frequency varied greatly from the desired behavior at the target frequency. In future research, we aim to improve our algorithm to automatically refine or discard such unintended solutions.



\end{appendix}